\documentclass{article}

\usepackage{arxiv}

\usepackage[utf8]{inputenc} 
\usepackage[T1]{fontenc}    
\usepackage{hyperref}       
\usepackage{url}            
\usepackage{booktabs}       
\usepackage{amsfonts}       
\usepackage{nicefrac}       
\usepackage{microtype}      
\usepackage{graphicx}
\usepackage{epstopdf, epsfig}
\usepackage{subfigure}
\usepackage{pgfplotstable}
\usepackage{booktabs}
\usepackage{filecontents}
\usepackage{longtable}
\usepackage{array}
\usepackage{amsmath} 
\usepackage{bm}
\usepackage{amsmath}%
\usepackage{MnSymbol}%
\usepackage{wasysym}%

\title{Experiments on wave propagation in grease ice: combined wave gauges and PIV measurements}
\author{
  Jean Rabault\\
  Department of Mathematics\\
  University of Oslo\\
  \texttt{jean.rblt@gmail.com} \\
   \And
  Graig Sutherland\\
  Environment and Climate Change Canada, Dorval, Canada\\
  and\\
  Department of Mathematics, University of Oslo\\
  \texttt{graigorys.sutherland@canada.ca} \\
   \And
  Atle Jensen\\
  Department of Mathematics\\
  University of Oslo\\
  \texttt{atlej@math.uio.no} \\
   \And
  Kai H Christensen \\
  Norwegian Meteorological Institute, Oslo\\
  \texttt{kaihc@met.no} \\
   \And
  Aleksey Marchenko \\
  The University Centre in Svalbard\\
  \texttt{Aleksey.Marchenko@unis.no} \\
}

\begin{document}
\maketitle

\begin{abstract}
Water wave attenuation by grease ice is a key mechanism for the polar regions, as waves in ice influence many phenomena such as ice drift, ice breaking, and ice formation. However, the models presented so far in the literature are limited in a number of regards, and more insights are required from either laboratory experiments or fieldwork for these models to be validated and improved. Unfortunately, performing detailed measurements of wave propagation in grease ice, either on the field or in the laboratory, is challenging. As a consequence, laboratory data are relatively scarce, and often consist of only a couple of wave elevation measurements along the length of the wave tank. We present combined measurements of wave elevation using an array of ultrasonic probes, and water kinematics using Particle Image Velocimetry (PIV), in a small-scale wave tank experiment. Experiments are performed over a wider frequency range than what has been previously investigated. The wave elevation measurements are used to compute the wave number and exponential damping coefficient. By contrast with a previous study in grease ice, we find that the wave number is consistent with the mass loading model, i.e. it increases compared with the open water case. Wave attenuation is compared with a series of one-layer models, and we show that they satisfactorily describe the viscous damping that is taking place. PIV data are also consistent with exponential wave amplitude attenuation, and a POD analysis reveals the existence of mean flows under the ice that are a consequence of the displacement and packing of the ice induced by the gradient in the wave-induced stress. Finally, we show that the dynamics of grease ice can generate eddy structures that inject eddy viscosity in the water under the grease ice, which would lead to enhanced mixing and participating in energy dissipation.
\end{abstract}

\keywords{Grease ice \and Water waves \and Attenuation \and Eddy structures \and PIV}

\section{Introduction}

Interactions between water waves and sea ice are a key mechanism for the polar regions, as they participate in shaping the extent and quality of sea ice covers. Better quantitative understanding of these interactions could lead to improved sea state and climate models and better predictions of sea ice hazards for human activities \citep{Pfirman1995129, KaiReport, Wadhams200998}. In recent years, there has been an increase in research into wave-ice-interaction motivated by both increased human activity in the polar regions, and new insights into some of the feedback mechanisms that may be involved in sea ice decline. In particular, it has been shown from satellite data that reduced sea-ice extent, by increasing the fetch available for wave development, has led to more energetic sea states in the Arctic basin, which in turn increases ice breakup and melting \citep{GRL:GRL51656}.

In nature, several types of sea ice are observed with increased distance from the free water limit as one progresses into the sea ice. Going from the open water into the ice-covered region, generally one first encounters ice in the form of grease ice slicks, pancake ice and broken ice floes of progressively increasing size. This region, the Marginal Ice Zone (MIZ), substantially attenuates the energy of the high frequency waves that would otherwise separate the inner continuous ice pack \citep{WeberArticle,TwoLayersModel,GRL:GRL52708}. Therefore, obtaining detailed understanding and quantification of the dominant mechanisms affecting wave propagation in the MIZ is critical to the description of polar regions. In the following, we will focus mostly on the interaction between waves and grease ice. Grease ice is composed of frazil ice crystals, typically disks of size 1 to 4~mm in diameter and 1 to 100~$\mu$m in thickness \citep{NewyearLab1}. Grease ice accumulates and forms slicks of typical thickness 10 to 20~cm in the arctic ocean \citep{Smedsrud2006171, 2011AnGla5277S}, that effectively damp high frequency waves, therefore, appearing visually similar to an oil slick \citep{NewyearLab1}.

Studying physical processes in the MIZ is a challenging task, since it is a region where sea ice comes in several different forms which has the consequence of making it difficult to describe theoretically and to model \citep{squire2016evolution,Collins2017dispersioninice}. The strong inhomogeneity of the ice in the MIZ makes it especially difficult to derive theoretical or numerical models with a basis in simple physical principles. In addition, field measurements are made challenging by the existence of a number of external factors, such as wind input, which can introduce artefacts in the data \citep{li2017rollover}. As a consequence, small scale experiments and simple analytical models have been used to investigate the effect of grease ice on wave propagation in the laboratory. Focusing on slush ice and pancake ice, \citet{Martin1981} were the first to perform laboratory experiments, allowing for a detailed investigation of the phenomena. Important conclusions reached from laboratory experiments included the fact that the mean thickness of the grease ice increases along the direction of the propagation of the waves, caused by the packing effect of the waves on the ice. Complex grease-ice dynamics that included global circulation of the ice under the influence of the waves were observed by the authors.

To our knowledge the first theoretical model describing wave attenuation by grease ice was presented by \citet{WeberArticle}. In this model, the ice is considered to be so viscous that it undergoes a `creeping motion', corresponding to a balance between pressure gradient and viscous stress. This should not be confused with ice creep \citep{wadhams1973attenuation}. This imposes a no-slip boundary condition under the ice, and all of the wave energy dissipation occurs in the underlying water. This is similar to the solution found by \citet{lamb1932hydrodynamics} (Equation 351.8) for waves propagating under an inextensible surface film, though the result of \citet{lamb1932hydrodynamics} should be converted to a spatial attenuation rate following the Gaster relation \citep{gaster_1962}. An effective eddy viscosity much higher than the molecular viscosity of water is required in the water layer for the model to be consistent with observations and laboratory experiments. While a crude simplification of reality, this simple model provides good agreement with both laboratory and field data when an empirically fit value for the effective viscosity of the water is used \citep{NewyearLab1, RabaultSutherlandGlaciology}. The use of a high effective viscosity is usually justified by the need to describe the existence of a large range of eddies under the ice that enhance mixing \citep{DeCarolis2002399, GRL:GRL53001}.

Next, \citet{NewyearLab1} introduced another one-layer model to compare with experimental results in a small wave tank facility. By contrast with the model of \citet{WeberArticle}, their model assumes that since most of the wave motion is usually concentrated near the free surface, attenuation can be approximated using an infinitely deep ice layer. This solution is similar to the calculation of \citet{lamb1932hydrodynamics} for waves propagating in a viscous fluid (Equation 349.21) with a no-stress boundary condition at the free surface. There also, the Gaster relation can be used for performing conversion between spatial and temporal attenuation rates. In the model of \citet{NewyearLab1}, the value of the viscosity to use in the ice layer is not known a priori, and is usually obtained from a fit to experimental data.

This model of \citet{NewyearLab1} was extended to a two-layer model by \citet{TwoLayersModel}, who considered the case of a finite ice thickness layer with inviscid water under the layer. This opened the way for the development of a variety of such two-layer models. \citet{DeCarolis2002399} formulated a model in which viscosity is also added to the water under the ice, while \citet{JGRC:JGRC11467} considered a model in which the water layer is inviscid, but the ice layer is viscoelastic (a Voigt model). Such models are a better description of reality since a clear separation between the grease ice and the water is observed in the experiments, and these two phases have very different mechanical properties. Better agreement is also observed between those models and laboratory data than with the previously mentioned one layer models \citep{NewyearLab2}, even though the estimation of model quality has often been based on curve fitting and visual impression rather than quantitative metrics (see, for a discussion of this issue, \citet{RabaultSutherlandGlaciology}). These models are usually applied indistinctly to grease ice, pancake ice, and continuous ice, as the parametrization they rely on can be tuned to accomodate for different ice rheologies.

However, two layer models are affected by at least two issues. First, while being a better description of reality, the parameters they rely on are obtained empirically from the experimental data rather than analysis of the underlying properties of the ice, and, as a consequence, they differ little from the arguably simplistic single layer models in this aspect. Therefore, it is difficult to know whether the better agreement observed with experimental data is due to a better description of the physics or is a mathematical artefact due to more fitting parameters being available. In addition, the model of \citet{Wang201090} has several issues that were raised by \citet{JGRC:JGRC21350}. In particular, the large number of roots in the dispersion relation makes it challenging to use such models for real-world applications, as selection of the physically relevant propagation mode is more challenging than it is for simpler models.

By contrast, a recently proposed alternative to these models \citep{SutherlandDissipation} assumes that, if the ice is thick and viscous enough, a portion of the ice consists of ``creeping motion'', and that the no-slip condition should be applied within the ice layer. If the ice is less thick or viscous, then this creeping motion does not exist and a solution similar to \citet{lamb1932hydrodynamics, NewyearLab1} can be used. This model has proven successful at describing a wide range of data, both from laboratory experiments and fieldwork studies \citep{SutherlandDissipation}. Moreover, it is mathematically straightforward to implement. Therefore, if further validated, it would appear as an interesting alternative for computing wave attenuation in the marginal ice zone.

Faced with the wide range of models presented in the literature, the natural reaction should be to test one or several of the theories against laboratory or field experiments. However, by contrast with the variety and complexity of the models presented in the literature, experimental data have remained scarce and indeed are limited to quite simple measurements such as single optical images \citep{Martin1981}, and the collection of a few single point measurements of wave elevation or exceeding pressure along one wave tank length \citep{Martin1981, NewyearLab1, Wang201090, Zhao201571}. Out of these publications, it should be noted that both the work of \citet{Wang201090} and \citet{Zhao201571} are investigating thin continuous ice covers rather than grease ice. This is probably at least partly due to the fact that measurements of waves in grease ice are challenging to perform: access to a carefully temperature-controlled wave tank is needed, and to grow ice in a specific fashion to obtain realistic grease ice before performing experiments is a process that can be time consuming on its own. Therefore, performing measurements requires both specific infrastructure and also more time and work than the acquisition of data on its own. In addition, owing to the constraints associated with growing ice at the surface of the wave tank, optical access can be challenging therefore making it more difficult to use modern measurement techniques such as Particle Image Velocimetry (PIV) and Particle Tracking Velocimetry (PTV). This is probably the reason why no study of wave propagation in grease ice published to this date has attempted PIV or PTV. While it is challenging to perform laboratory measurements, the situation is even worse regarding field data. A combination of factors, such as the ice condition (that cannot be controlled), the difficulties associated with working in polar environments, and external sources of disturbance (e.g. currents, wind, ship wakes) make it difficult to obtain more than single point wave elevation measurements when working in the field. While field data have proven valuable \citep{WeberArticle, JGRC:JGRC4212, SquireOOWASI, JGRC:JGRC21649, RabaultSutherlandGlaciology}, they may not be sufficient to elucidate the mechanisms at stake, therefore emphasizing the need for more laboratory data. In addition, it should be underlined here that \citet{RabaultSutherlandGlaciology} study specifically grease ice, while \citet{WeberArticle}, \citet{JGRC:JGRC4212} and \citet{SquireOOWASI} deal with the MIZ, which contains a significant amount of grease ice, but not exclusively.

In this article, we present experimental data about wave propagation in grease ice, both traditional wave elevation measurements using an array of ultrasonic gauges, and also for the first time direct measurements of the water dynamics using PIV. In addition, we perform our wave elevation measurements over a wider frequency range than what is usually reported in the literature. The organization of the paper is as follows. First, we present the methodology used for analyzing the data obtained in the present study and comparing it with data obtained in previous studies. Both the processing of the wave elevation and the PIV data are detailed there. Second, we present the results obtained from processing of the data. Finally, we discuss the results obtained with regards to real world applicability.

\section{Laboratory measurements: methodology}

Two sources of laboratory data are used in all the following: the data obtained in the present study and the data reported by \citet{NewyearLab1}, which are available from Tables 1 and 2 of their article, comprising two experiments in the same test facilities. They report experimental details in the corresponding article, and the data were used in one later study \citep{NewyearLab2}. \citet{NewyearLab1} used a flap-type paddle (i.e., a plate rotating around its lower edge) to generate waves in a wave tank $3.5$~m long, $1$~m wide and $1$~m deep, which was filled with a depth $H = 50$~cm of water. The wave tank was located in a cold room, which allowed the authors to create slush ice in a controlled environment. The slush ice thickness reported by the authors was $11.3$ and $14.6$~cm for experiments 1 and 2, respectively. The frequency range investigated by \citet{NewyearLab1} is relatively small, between $1.0$ and $1.6$~Hz, and the lower end of the frequency range does not strictly fulfill the deep water waves condition, which adds complexity to the situation in addition to the effect of grease ice.

The data collected in the present study were obtained in a cold room facility located at the University Center in Svalbard (UNIS). All the details of the actuation and logging system are released as open source material, see Appendix A for more details and a link to the description of the system and code. A custom-designed flat-blade paddle (i.e., the whole plate moves back and forth while staying vertical) was used to generate waves over a frequency range between $1.5$ and $2.7$~Hz, with an increment of $0.1$~Hz. This is a wider frequency range than investigated previously by \citet{NewyearLab1}. Measurements were performed for 6 different wave-paddle displacement amplitudes at each of those frequencies, for a total of 78 measurements. The wave tank is built in transparent acrylic with the dimensions $3.5$~m long, $0.3$~m wide and $0.5$~m high, and the water depth is $25$~cm. A damping beach is located at the opposite end of the wave tank to the paddle.

During the experiments, the effect of the grease ice alone absorbs most of the wave energy. As a consequence, very little water motion is still present at the position of the beginning of the damping beach, and the water at the end of the damping beach is essentially still. Consequently, we do not fear that energy may be reflected at the end of the wave tank. Removable insulation foam is placed under and on the sides of the wave tank in order to avoid ice formation on the wave tank walls during grease ice growth, and can be taken away for gaining optical access when PIV measurements are performed. The slush ice is generated once at the beginning of the series of measurements by setting the temperature of the cold room to $-6^{\circ}$C, and turning the paddle on so that waves of frequency $2$~Hz are generated with a large paddle motion amplitude. As a consequence, breaking waves are generated that introduce turbulence in the whole wave tank creating the right conditions for grease ice formation. In addition, small ice crystals generated in a slush-ice machine are added in the water to speed up the process. As a result, the grease ice layer obtained is about $4$~cm thick when the water is at rest and the ice spreads evenly in the whole wave-tank. The methodology used for generating the grease ice is similar to the one reported by \citet{NewyearLab1}. Following ice formation, the room temperature was set to 0 degrees so that no significant ice melting nor freezing took place.

A total of 6 ultrasonic gauges (model U-gauge S18U, from Banner Engineering) are used to measure wave elevation. The gauges are located $50$, $66.6$, $88$, $110.5$, $133.9$, and $179.9$~cm away from the wave-maker. The distance between consecutive gauges is deliberately varied, as this helps detecting possible aliasing effects when performing cross-correlation analysis of the wave elevation signal. The gauges have a resolution of $0.5$~mm, and perform measurements at a frequency of $100$~Hz. The noise level of gauges 4 and 6 is much higher than for the other gauges, possibly because of ice formation on the gauges, and therefore we only use the signals from gauges 1, 2, 3, and 5.

The data from the ultrasonic gauges are used in two ways. For each gauge, the wave amplitude is computed by integrating the Fourier spectrum of wave elevation around the peak frequency. The Fourier spectrum is computed from time series of duration 300 seconds using a $50\%$ percent overlap on $40$ seconds time windows, and the width of the integration domain around the peak frequency is $1.0$~Hz. For each run, the wave amplitudes obtained from the different gauges are then fitted to a decreasing exponential in order to extract the coefficient of wave damping, using:

\begin{equation}
\frac{\partial a}{\partial x} = - \alpha a,
\end{equation}

\noindent where $a$ is the wave amplitude in the convention that $\eta(t) = a \cos(\omega t)$ is the wave elevation at a fixed point in space, with $\omega = 2 \pi f$ the wave angular frequency in $rad / s$, $f$ the wave frequency, $x$ the distance to the wave-maker, and $\alpha$ the spatial decay coefficient that describes wave attenuation. This processing method is similar to what was presented by \citet{Sutherland201788}.

The spatial damping coefficient obtained experimentally, $\alpha$, is compared with the result obtained analytically in Eqn. (4.15) of \citet{WeberArticle}, the no-stress boundary condition of \citet{lamb1932hydrodynamics} and \citet{NewyearLab1}, and both Eqn. (10) and Eqn. (13) of \citet{SutherlandDissipation}. More explicitly, Eqn. (4.15) of \citet{WeberArticle} corresponds to the wave dissipation due to the presence of an inextensible surface cover \citep{lamb1932hydrodynamics} and can be written as:

\begin{equation}
  \alpha_{in} = \frac{1}{2} \nu \gamma k / c_g,
  \label{weber_equation}
\end{equation}

\noindent where $\nu$ is the kinematic eddy viscosity, which can be taken higher than the viscosity of water to account for eddies and, in some measure, grease ice properties, $\gamma =  \sqrt{\omega / 2 \nu}$ is the inverse thickness of the surface boundary layer, $k$ is the wavenumber, and $c_g$ is the group velocity.

The damping obtained in the case of a no-stress boundary condition \citep{lamb1932hydrodynamics,NewyearLab1} can be written as:

\begin{equation}
  \alpha_{ns} = 2 \nu k^2 / c_g,
  \label{damping_NM97}
\end{equation}

\noindent where $\nu$ is again a fitting parameter used to model grease ice properties.

Finally, Eqn. (10) of \citet{SutherlandDissipation}, can be written as:

\begin{equation}
  \alpha_{\epsilon} = \frac{1}{2} \epsilon h k^2,
  \label{sutherland_damping_v1}
\end{equation}

\noindent where $0 < \epsilon < 1$ is the fractional ice thickness, and $h$ is the total ice thickness. A fractional ice thickness $\epsilon = 0.7$ was previously found to model well a wide range of experimental datasets \citep{SutherlandDissipation}. Eqn. (13) of \citet{SutherlandDissipation} can be written as:

\begin{equation}
  \alpha_t = 2 h^2 k^3.
  \label{sutherland_damping_v2}
\end{equation}

The difference between Eqn. (\ref{sutherland_damping_v1}) and Eqn. (\ref{sutherland_damping_v2}), which correspond respectively with equations (10) and (13) of the model by \citet{SutherlandDissipation}, is that in the former case the grease ice is considered thick enough so that the top of the ice is still.

Since the wave-tank is of finite width and depth, damping is introduced by boundary layers developing on the side and bottom walls. Those sources of damping must be assessed to confirm that the main damping effect observed arises from the grease ice. The damping introduced by the laminar boundary layers created by the wave motion can be written as \citep{van1966boundary, Sutherland201788}:

\begin{equation}
  \alpha_{bs} = \nu_{0} \gamma_0 k \left( \frac{1}{\sinh(2kH)} + \frac{1}{kB}\right) / c_g,
  \label{DampingBL}
\end{equation}

\noindent where $\nu_{0}$ is the kinematic viscosity of water, $\gamma_0 = \sqrt{\omega / 2 \nu_{0}}$ is the inverse boundary thickness, and $B$ is the width of the wave tank. The damping coefficient $\alpha_{bs}$ is found to be two orders of magnitude smaller than the damping measured experimentally, and therefore is neglected in all of the following.

The wave number is obtained from performing a cross-correlation analysis of adjacent ultrasonic gauges. For this, the phase difference between waves recorded by the gauges $m$ and $n$, $\phi_{mn}$, is obtained from the co-spectral density between adjacent sensors, $S_{mn}$, as:

\begin{equation}
  \phi_{mn} = \tan^{-1} \left( \frac{\Im[S_{mn}(f)]}{\Re[S_{mn}(f)]} \right),
  \label{get_wave_phase_shift}
\end{equation}

\noindent where $\Im$ and $\Re$ indicate the imaginary and real part, respectively. The co-spectral density is computed using windows of length 41 s, with a 50\% percent overlap. The equation describing the propagation of waves phase can be written as:

\begin{equation}
  \phi_{mn} = \bm{k} \cdot \bm{x}_{mn},
  \label{get_wave_vector}
\end{equation}

\noindent where $\bm{k}$ is the wave vector and $\bm{x}_{mn}$ the distance between gauges $m$ and $n$. The wave vector corresponding to the frequency generated by the paddle can therefore be obtained from Eqn. (\ref{get_wave_vector}) since the distance between the gauges is known and the phase shift between adjacent gauges can be computed using Eqn. (\ref{get_wave_phase_shift}). This methodology is similar to what was presented by \citet{JGRC:JGRC21649} and by \citet{marchenko2017field} in the case of field data. Results are compared with both the finite depth, open water dispersion relation and the mass loading effect. The open water dispersion relation is written as:

\begin{equation}
  \omega^2 = g k \tanh \left( kH \right),
  \label{free_water_dispersion_relation}
\end{equation}

\noindent with $H$ water depth. The dispersion relation including mass loading is taken from \citet{NewyearLab1}:

\begin{equation}
  \omega^2 = \frac{gk \rho_{water} \tanh \left( kH \right)}{\rho_{water} + c \rho_{ice} h k \tanh \left( kH \right)}, \label{mass_loading_dispersion_relation}
\end{equation}

\noindent with $c \approx 0.5$ the volumetric fraction of ice.

When analyzing laboratory data, we consider that dissipation is produced by the viscous effects introduced in water and / or the ice, as calculated by \citet{WeberArticle}, \citet{lamb1932hydrodynamics}, or \citet{SutherlandDissipation}, while modification of the wave number is mostly due to the mass loading of the grease ice layer, and that as a first approximation both effects can be accounted for separately.

\begin{figure}
  \begin{center}
    \includegraphics[width=.45\textwidth]{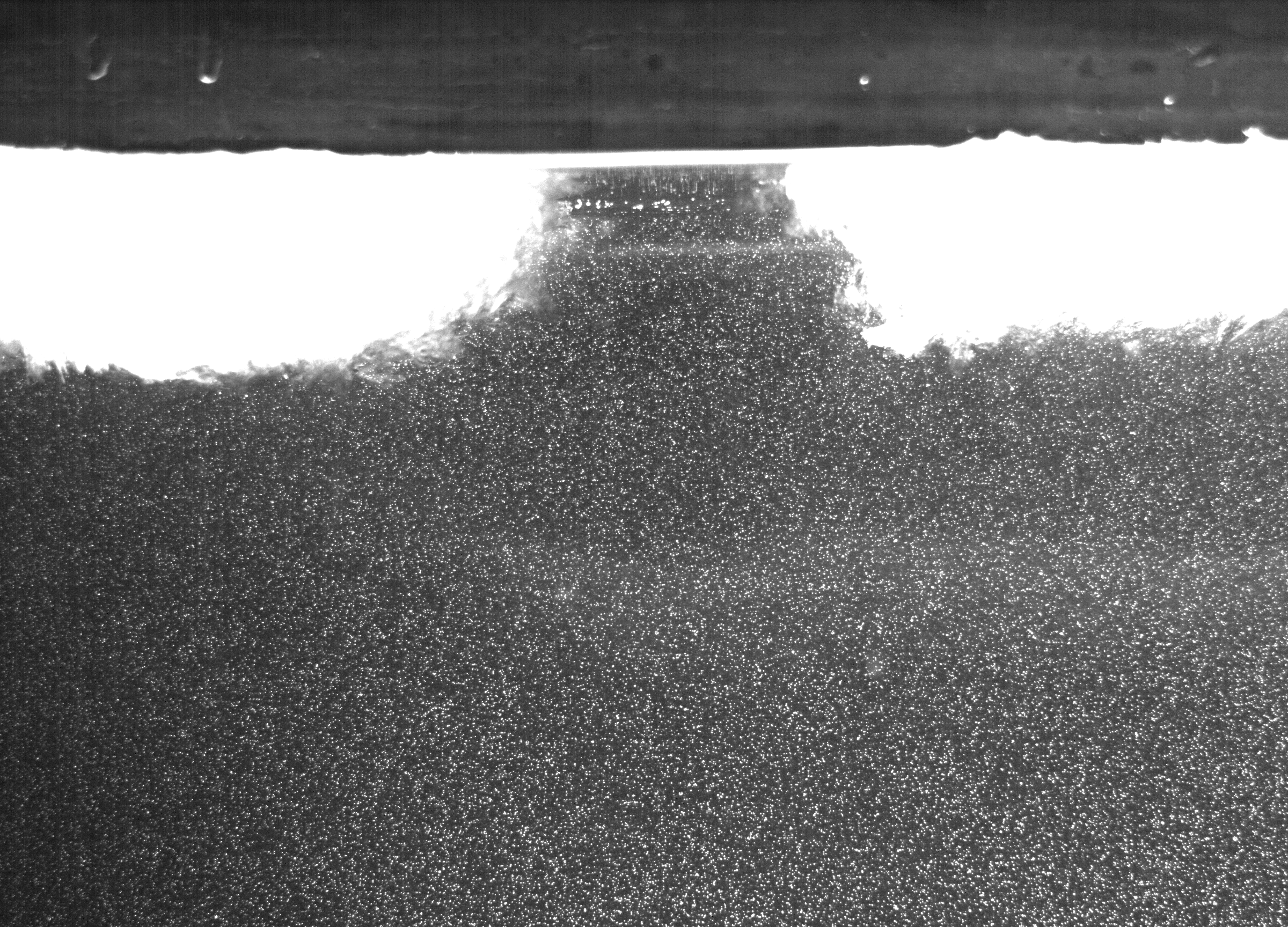}
    \includegraphics[width=.45\textwidth]{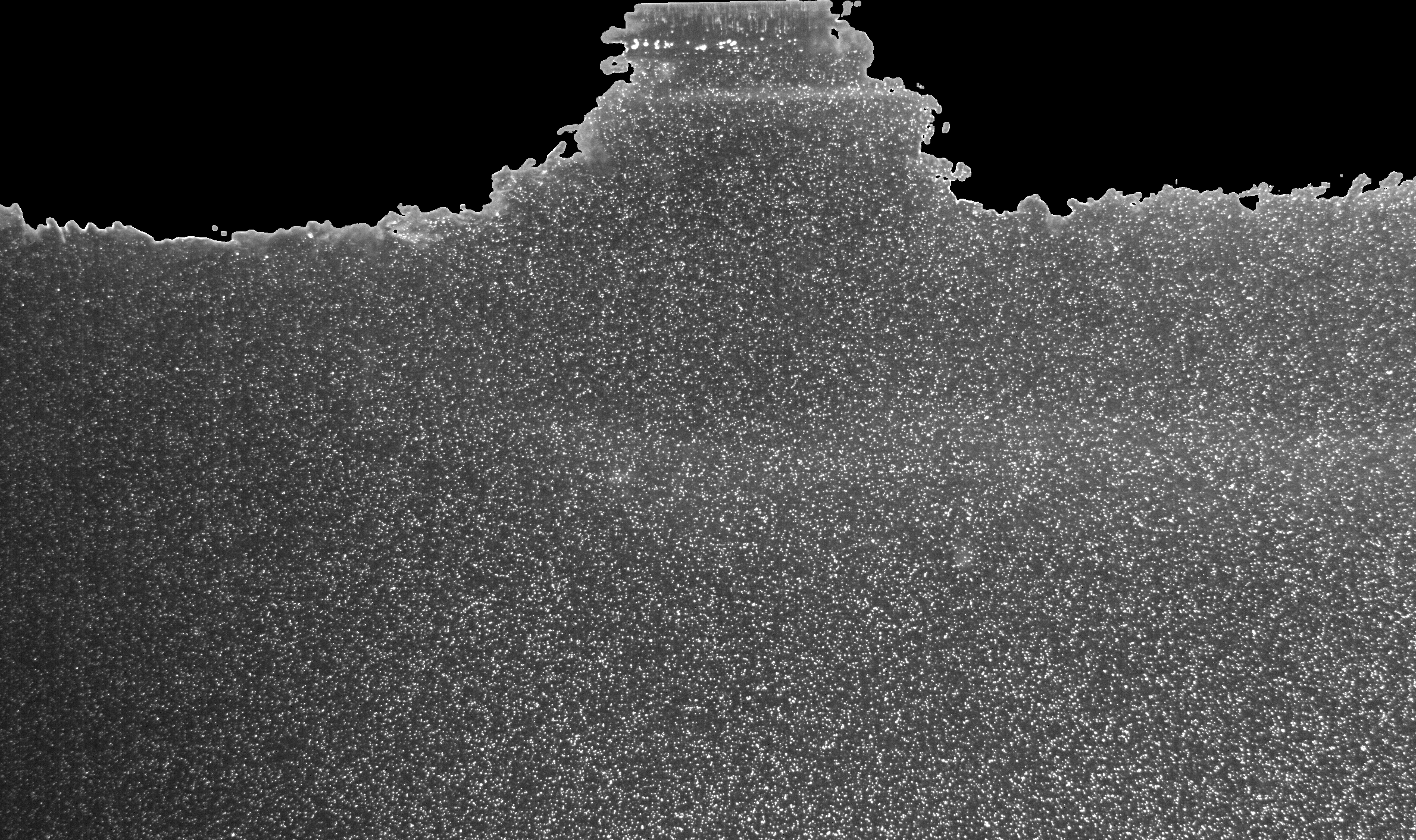}
    \caption{\label{filtering} Picture extracted from the case 2, which corresponds to an initially discontinuous ice cover. Left: raw picture before filtering. Both the water surface and the region occupied by grease ice are clearly visible. Right: picture after filtering.}
  \end{center}
\end{figure}

When performing PIV, a white color linear LED array is used for illumination. Spherical Polyamid Seeding Particles (PSP) of diameter $50$~$\mu$m are used for seeding. A single Falcon2 4M camera is used for taking pictures at a frame rate of 75Hz. This PIV setup is similar to that used previously by authors of this paper \citep{rabault2016ptv}. The measurement plane is located in the middle of the wave-tank, along the main direction of the wave-tank and wave propagation, and a mirror inclined $45^{\circ}$ is used for finely adjusting the position of the light sheet. Filtering is used to detect both the free water surface and the area that corresponds to grease ice, as shown in Fig. \ref{filtering}. Consecutive frames are used for performing PIV using the Digiflow software \citep{Digiflow}.

The velocity fields obtained from PIV are processed using the Proper Orthogonal Decomposition (POD). POD is based on the singular value decomposition (or an eigenvalue analysis of the positive semi-definite matrix $X^{T}X$, where $X$ is the snapshot matrix, see under), and is known under different names depending on the field of study, such as 'Principal Component Analysis' (from statistics) and 'Empirical Decomposition Functions' (from meteorology). In the present case, the POD is computed in Matlab from the Singular Value Decomposition (SVD) of the snapshots matrix (for detailed explanations about POD and the snapshot method, see for example \citet{berkooz1993proper, Kerschen2005}). The snapshot matrix, $X$, is constructed as:

\begin{equation}
  X =
  \begin{bmatrix}
    u_1^1       & ... & u_n^1  \\
    \vdots      & ... & \vdots \\
    u_1^k       & ... & u_n^k \\
\end{bmatrix},
\end{equation}

\noindent where $u_i^j = u(\bm{x}_j, t_i)$, with $\bm{x}_j$ the position of the point considered, and $t_i$ the time of the snapshot. Both the $X$ and $Y$ components of velocity are stored in $u$, by letting the $u_i^{1..k/2}$ represent the $X$ component and the $u_i^{k/2+1..k}$ represent the $Y$ component, respectively. As a consequence, each column of $X$ contains the 2D, two component velocity field at a given time reshaped into a 1D vector. The SVD decomposition is then computed as:

\begin{equation}
  X = U S V^*,
\end{equation}

\noindent where $U$ and $V$ are unitary matrices. The diagonal coefficients of $S^2$ give the energy of each mode, while $U$ and $V$ contain the POD modes and POD mode coefficients. The POD modes together with the POD mode coefficients contain the full information about the velocity field at all times. In particular, the description of the velocity field obtained with POD is interesting as it optimizes the energy content in the modes of lower index, therefore effectively extracting an ordered list of the most energetic coherent structures of the flow. Therefore, the POD can be applied to produce a condensed overview of the main structures present in the data of a dataset too large to be analyzed by simple visualization techniques.

\section{Results}

\subsection{Ultrasonic gauges and comparison with models for damping}

\begin{figure}
  \begin{center}
    \includegraphics{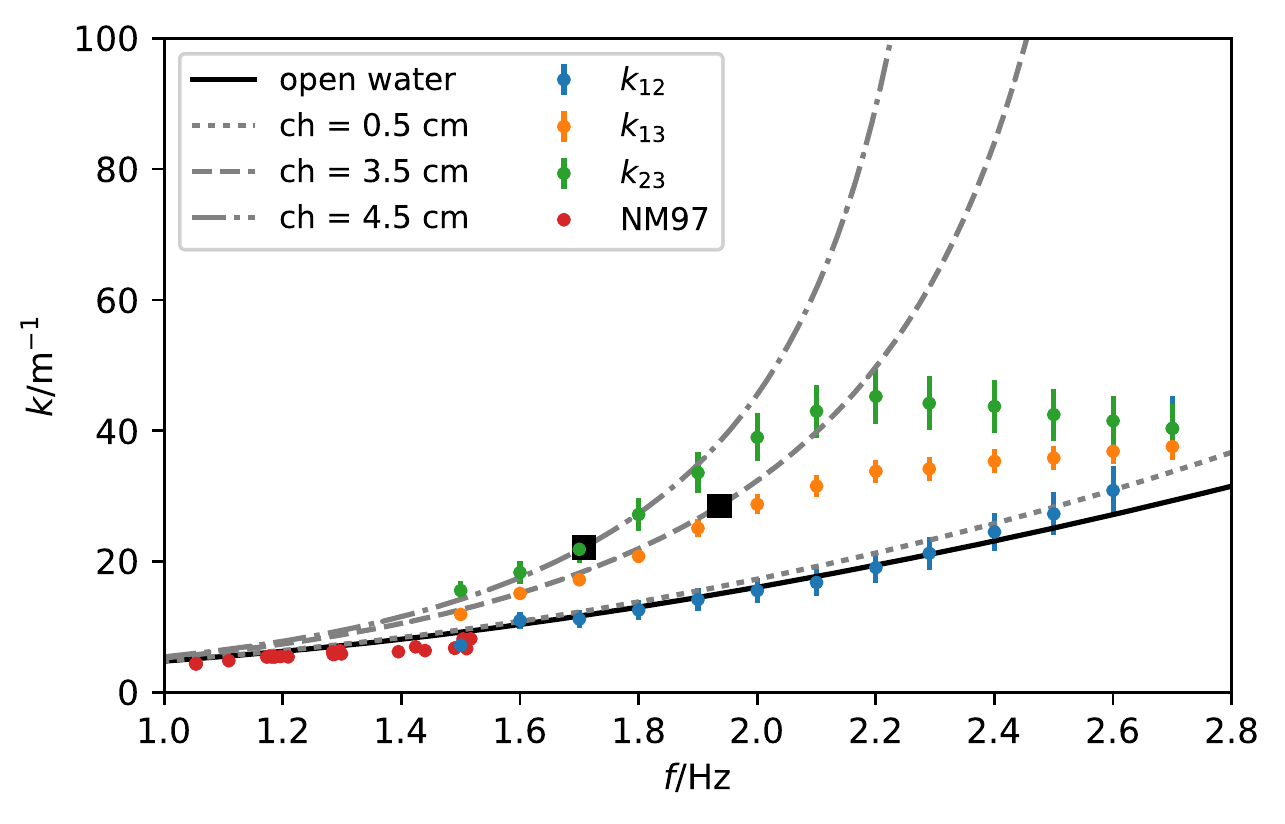}
	  \caption{\label{wavelength} Wavenumber reported by \citet{NewyearLab1} (noted NM97 in the legend), and obtained in the present study using Eqn. (\ref{get_wave_vector}) for three different pairs of gauges (1 and 2, 1 and 3, and 2 and 3 are denoted as $k_{12}$, $k_{13}$ and $k_{23}$, respectively). The black line indicates the open water dispersion relation. Dotted lines indicate the mass loading dispersion relation, for different ice thicknesses. Large black dots ( $\blacksquare$ ) denote $k = (ch)^{-1}$, in the case of each ice thickness. The mean values and 95\% confidence intervals are computed for each frequency based on the experiments performed at different wave amplitude.}
  \end{center}
\end{figure}

The wavenumbers reported by \citet{NewyearLab1}, together with the results obtained in our experiments using the cross-spectrum analysis and Eqn. (\ref{get_wave_vector}), are presented in Fig. \ref{wavelength}. Data obtained from the experiments 1 and 2 of \citet{NewyearLab1} are indicated by red dots, while our data are indicated by dots of varying color, which depends on the pair of gauges used in the cross correlation analysis. In the case of the data from our experiments, since wave amplitude is found to have no significant effect on the values obtained the mean values and 95\% confidence intervals are computed based on the results obtained at the different wave amplitudes. The open water dispersion relation curve is computed from Eqn. (\ref{free_water_dispersion_relation}), using the water depth corresponding to the experiments performed by the present authors. Owing to the scaling of the figure, the difference with the open water dispersion relation corresponding to the water depth reported by \citet{NewyearLab1} (not shown) is barely visible. The mass-loading dispersion relations, corresponding to different grease ice thicknesses, are computed from Eqn. (\ref{mass_loading_dispersion_relation}) using also the water depth corresponding to the experiments performed by the present authors.

Several observations can be made from Fig. \ref{wavelength}. Firstly, while the data from \citet{NewyearLab1} could indicate a limited deviation from the open water dispersion relation in the opposite direction to the mass loading effect, our experiments performed over a wider frequency range show the opposite effect. There are several possible sources to this discrepancy. A first explanation could be that the properties of the grease ice generated may be different between both sets of experiments. However, this does not seem likely as the methodology used for generating grease ice was similar in both cases. Another explanation could be that the small deviation from the open water dispersion relation reported by \citet{NewyearLab1} arises from sensor noise, random experimental error, or is of other technical origin. Finally, the different grease ice thicknesses used by \citet{NewyearLab1} and the present authors could also play a role in this discrepancy. Whatever the explanation, Fig. \ref{wavelength} underlines the importance of performing laboratory measurements of wave propagation in ice on a frequency range as wide as possible, in order to observe large deviations from the open water baseline.

Secondly, the dispersion relation obtained in our experiments is different depending on the pair of gauges which is used for computing the wave number. The origin of this difference comes from the varying mean thickness of the grease ice layer under the influence of the wave-induced stress gradient, as was reported already by \citet{Martin1981}. In addition, changes in the concentration $c$ of the grease ice (which, unfortunately, was not measured in this experimental setup) may participate in the phenomenon, as $ch$ is the relevant parameter deciding the shape of the dispersion relation predicted by the mass loading model. Between gauges 1 and 2, which are closest to the wave paddle, the grease ice layer is very thin and the dispersion relation corresponds well to the mass loading effect obtained with an effective grease-ice layer thickness of $0.5$~cm. Between gauges 2 and 3, that are further away from the wave paddle, an effective grease ice thickness $ch = 4.5$~cm gives the best agreement with experimental data, which corresponds to a true physical grease ice thickness of $9$~cm. This is in good agreement with the initial measurement of the grease ice thickness, which was around $4$~cm when water was at rest before any piling of the ice took place. Using signals from gauges 1 and 3 for computing the wave number leads to an intermediate result, which is a weighted averaged of both values.

Thirdly, the mass-loading dispersion relation accurately describes the wave number obtained from experiments up to the point where $k= (ch)^{-1}$, which is indicated by a black square in Fig. \ref{wavelength} for each ice thickness. Above this frequency, the wave number obtained experimentally deviates from the value predicted by the mass-loading model, and reaches a plateau or even decreases slightly. This effect does not have its origin in aliasing of the phase shift between the gauges, as the corresponding aliasing is expected for higher frequencies than those occurring at the start of this deviation, and is accounted for in the processing. As a consequence, the deviation from the mass loading model is most likely due to the fact that some of the assumptions at the origin of this model are not verified at higher frequencies. One such assumption, which fails above $k = (ch)^{-1}$, is that the ice as a thin, heavy layer on top of the water column.

\begin{figure}
  \begin{center}
    \includegraphics{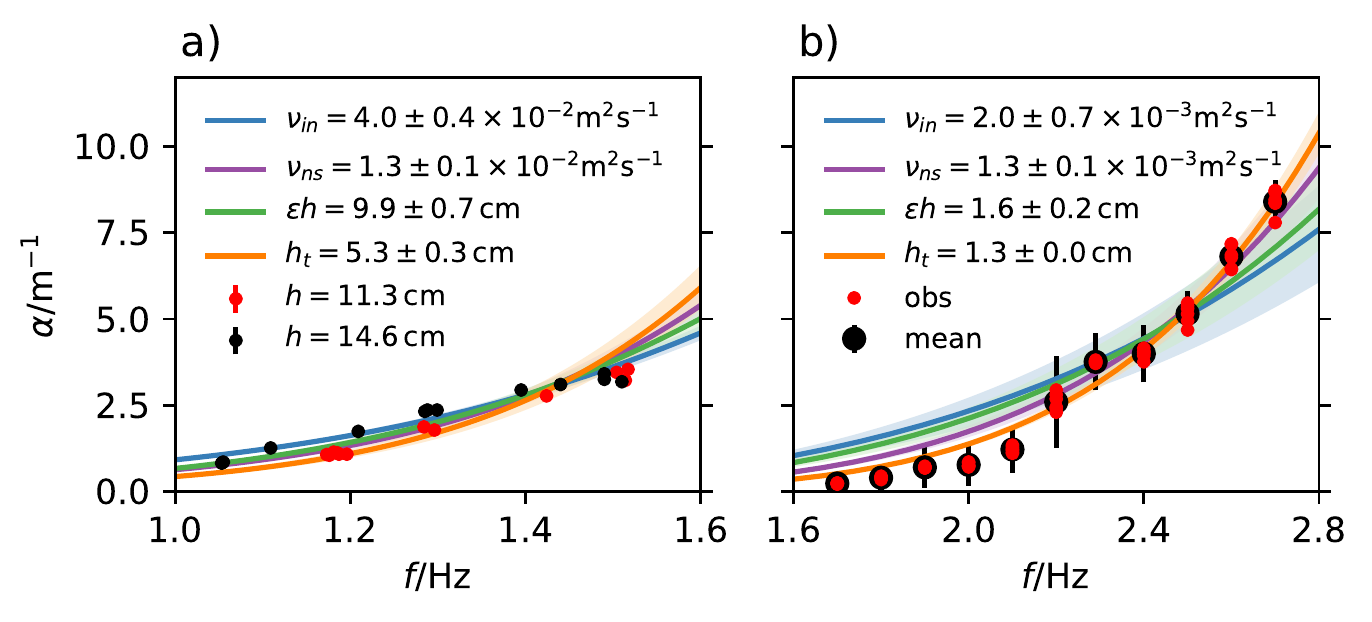}
	  \caption{\label{fig_damping} Spatial damping coefficient $\alpha$ as a function of wave frequency $f$. Blue curves indicate the results obtained with the one layer model of \citet{WeberArticle} from Eqn. (\ref{weber_equation}), magenta curves indicate the results obtained with the one layer model of \citet{lamb1932hydrodynamics} from Eqn. (\ref{damping_NM97}), green curves indicate the results obtained with the thick ice model of \citet{SutherlandDissipation} from Eqn. (\ref{sutherland_damping_v1}), and orange curves with the thin ice model of \citet{SutherlandDissipation} from Eqn. (\ref{sutherland_damping_v2}). The shaded region denotes 1 standard deviation in the best fit parameters. The attenuation measured at each frequency is computed as the average of all experiments performed for different wave amplitudes, and the error bars correspond to the 95\% confidence intervals. Left: results with the data reported by \citet{NewyearLab1}. Right: results with the data obtained by the authors.}
  \end{center}
\end{figure}

Results for wave damping are presented in Fig. \ref{fig_damping}. The theoretical curves for the spatial damping coefficient are indicated in different colors depending on the model used. Blue curves indicate the results obtained from the best non-linear fit with Eqn. (\ref{weber_equation}), magenta curves indicated the result obtained from the best non linear fit with Eqn. (\ref{damping_NM97}), green curves indicate the results obtained from Eqn. (\ref{sutherland_damping_v1}), and orange curves with Eqn. (\ref{sutherland_damping_v2}). Results are presented for both the data of \citet{NewyearLab1} and the data obtained by the present authors. Due to the limited frequency range investigated, little difference is observed between the quality of the fit of the different models to the data from \citet{NewyearLab1}.

In contrast, large differences between the quality of the fit for each of the models are observed in the case of the newly collected data. The value for the damping coefficient at each frequency is found to be predominantly independent of wave amplitude, and, therefore, all results presented are averaged over all amplitudes for a given frequency. This is clearly visible from the size of the error bars on Fig. \ref{fig_damping}, and from the raw data (made available in Appendix B, Table \ref{all_data_table}). We also provide an additional illustration of this fact in Appendix C, Fig. \ref{non_dependence_alpha}. It is clear that the use of Eqn. (\ref{sutherland_damping_v2}) gives the best agreement to the experimental data, while Eqn. (\ref{weber_equation}) performs the worst. The statistics characterizing the quality of fit for each model with the data collected in this paper are summarized in Table \ref{table_quality_fit}, and confirm the visual impression of Fig. \ref{fig_damping}. In particular, the coefficient of determination $R^2$ \citep{rao73} increases from $0.83$ to $0.98$ when using Eqn. (\ref{sutherland_damping_v2}), compared with Eqn. (\ref{weber_equation}). In addition, both the Mean Absolute Error (MAE) and Root Mean Square Error (RMSE) are reduced by a factor of about $3$ and a factor of about $1.5$ respectively, between the results predicted using Eqns. (\ref{damping_NM97}) and (\ref{sutherland_damping_v2}). As expected, Eqn. (\ref{sutherland_damping_v1}) performs better than the model of \citet{WeberArticle}, but is less applicable to ice of such low thickness than Eqn. (\ref{sutherland_damping_v2}). The effective ice thickness corresponding to the best fit to both versions of the model from \citet{SutherlandDissipation} should be expected to reflect a weighted average of the grease ice thickness along the region where damping is happening, which is confirmed by comparing the results in Fig. \ref{wavelength} and \ref{fig_damping}. Similarly to what was observed under the analysis of the wavenumber data, we find that wave damping should be analyzed on a frequency range as wide as possible in order to be able to clearly identify the main trends in the wave attenuation. In particular, the transition in the damping coefficient relatively to Eqn. (\ref{weber_equation}), at around $2.2$~Hz, is made clearly visible by the data collected in the present study, while it was not visible from the data collected by \citet{NewyearLab1}.

\begin{table}
  \begin{center}
    \begin{tabular}{lccc}
      \hline
      Model  & $R^2$ & MAE & RMSE \\
      \hline\hline
      One layer inextensible surface \citep{WeberArticle} & 0.83 & 1.12 & 0.98 \\
      \hline
      One layer stress-free surface \citep{NewyearLab1} &  0.95  & 0.60  &  0.55 \\
      \hline
      Model 1, Eqn. (10) of \citet{SutherlandDissipation} & 0.89 & 0.90 & 0.80 \\
      \hline
      Model 2, Eqn. (13) of \citet{SutherlandDissipation} & 0.98 & 0.39 & 0.34 \\
      \hline
    \end{tabular}
    \caption{Coefficient of determination ($R^2$), Mean Absolute Error (MAE) and Root Mean Square Error (RMSE) of the best fit between the data collected by the present authors and both the one layer models of \citet{WeberArticle}, \citet{NewyearLab1}, and both versions of the model of \citet{SutherlandDissipation}.}
    \label{table_quality_fit}
  \end{center}
\end{table}

In a recent paper, \citet{meylan2018dispersion} suggested that power laws may provide an insight inside the dissipation mechanisms at play for waves in ice. There are several issues to a power law approach, in particular the fact that the frequency ranges considered are often far from covering even a couple of decades and therefore considerable uncertainties may be introduced by experimental errors. In addition, a number of additional mechanisms, such as wind input \citep{li2017rollover}, may influence the fitting of power laws on experimental data. When applying a power law least-square fitting to our data, we obtain an exponent of $6.5 \pm 0.5$ (considering a 3-$\sigma$ confidence interval), in good agreement with the prediction of Eqn. (\ref{sutherland_damping_v2}) considering the sources of uncertainty previously mentioned (as this is peripheral to the main point of the paper, the fit is presented in Appendix C, Fig. \ref{alpha_power}). This is in contrast with the results of \citet{meylan2018dispersion}, and may indicate that different regimes are being considered. In particular, the power law of \citet{meylan2018dispersion} corresponds to weakly attenuated waves propagating for a long distance in the MIZ and the packed ice, while our experiments rather describe the sharp attenuation of high frequency waves by a highly viscous grease ice layer. Also, the data from \citet{meylan2018dispersion} may include non-linear effects, and involve a combination of different kinds of ice which is not the case in our experiment.

A summary of the processed data collected for this study is summarized in Table \ref{all_data_table} (see Appendix B).


\subsection{Measurements of kinematics}

Owing to the large amount of data generated when performing PIV, image processing was restricted to two cases. Both cases correspond to a wave frequency of $2.5$~Hz, and the same amplitude for the motion of the paddle. The corresponding wave steepness in the open water is $\epsilon = k a \approx 0.12$. In the first case, a continuous layer of slush ice was present since the beginning of the experiment and got piled up by the wave-induced stress, but no large slush-ice motion was present. In this case, the field of view of the camera is looking at the water motion under the piled grease ice. In the second case, an initially discontinuous layer of slush ice was present, and the initially separate slush-ice packs were accelerated by the waves and collided in the area of the field of view of the camera before piling up. Therefore, the first situation is ideal for testing theoretical models while the second situation is dynamic and may be more representative of turbulent conditions obtained in the ocean.

In both cases, the POD snapshot method presented in the Methodology section is used. This allows for the extraction of information from all the PIV frames into a reduced number of high energy modes, and, therefore, makes analysis of the data easier and less sensitive to noise. The decay of the energy content of the POD modes is very sharp, with both cases containing around 87 and 80 percent of the total energy in the first two modes respectively. This confirms that the POD is able to extract significant flow features.

\begin{figure}
  \begin{center}
    \includegraphics[width=.49\textwidth]{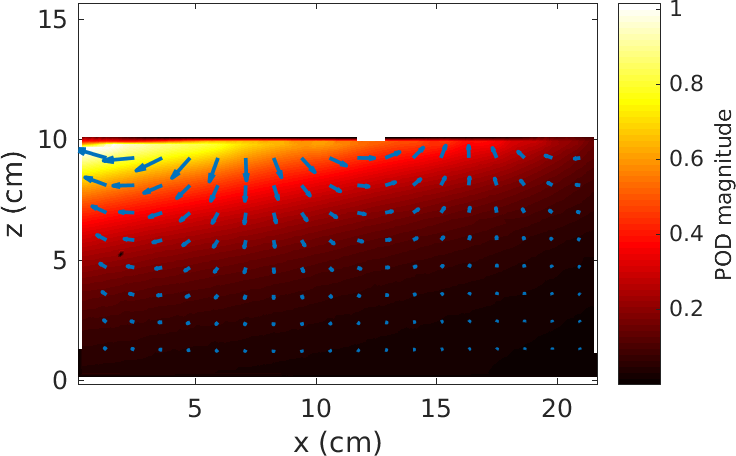}
    \includegraphics[width=.49\textwidth]{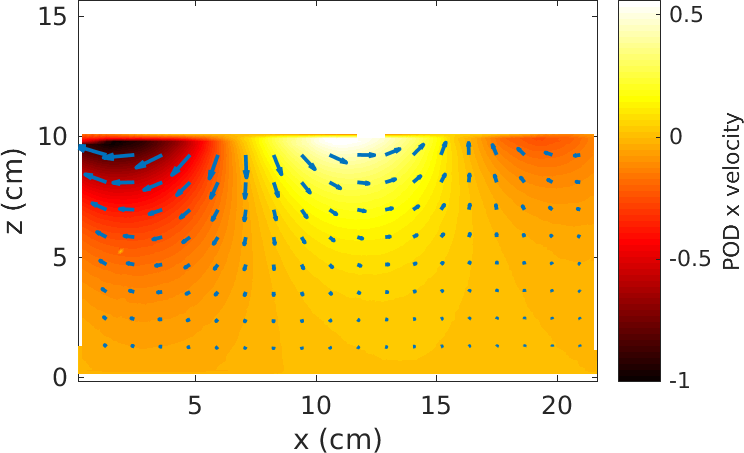}
    \includegraphics[width=.49\textwidth]{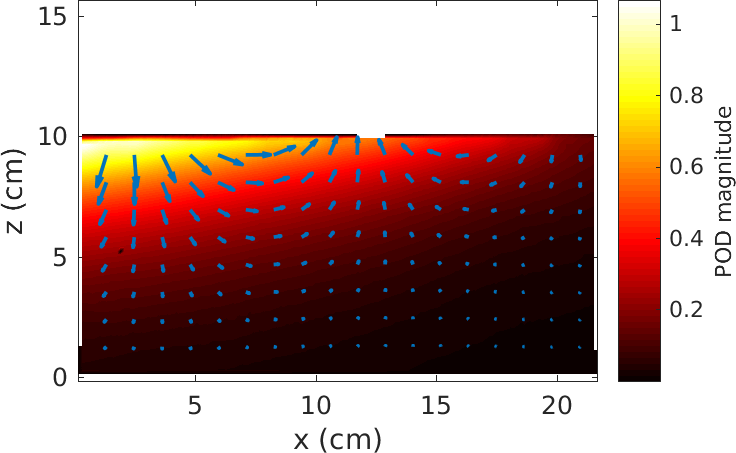}
    \includegraphics[width=.49\textwidth]{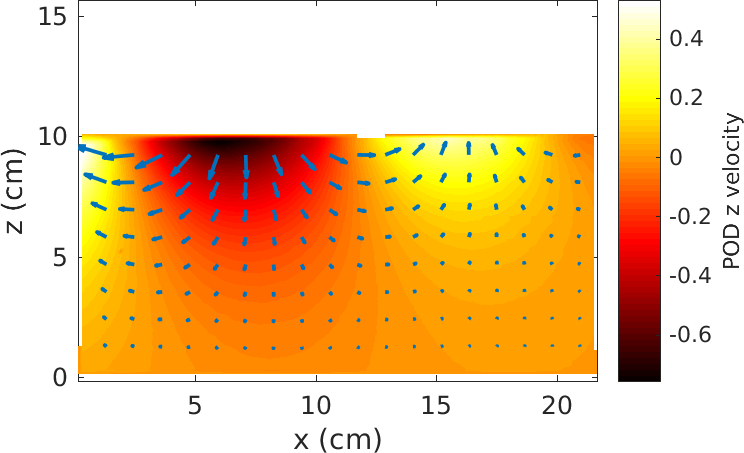}
    \caption{\label{case1_PIV_continuous_mode12} Summary of the POD modes 1 and 2, obtained in case 1. Left: flow direction (arrows, scaled by flow magnitude) and magnitude (colormap) for modes 1 (top) and 2 (bottom). Right: flow direction (arrows, scaled by flow magnitude) together with the $X$ (top) and $Y$ (bottom) velocity fields (colormap), for mode 1.}
  \end{center}
\end{figure}

\begin{figure}
  \begin{center}
    \includegraphics[width=.49\textwidth]{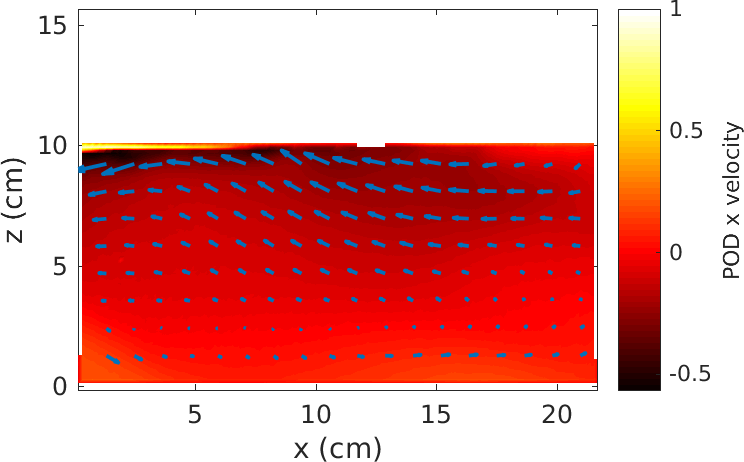}
    \includegraphics[width=.49\textwidth]{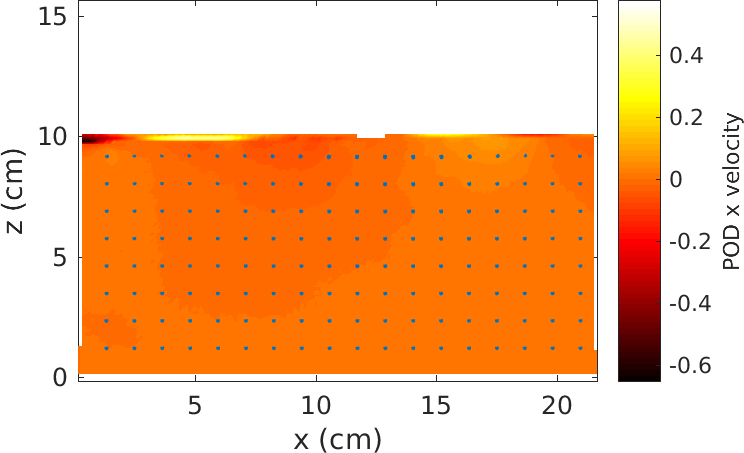}
    \caption{\label{case1_PIV_continuous_mode34} Summary of the POD modes 3 and 4, obtained in case 1. Left: POD mode 3. Right: POD mode 4. Arrows indicate flow direction, and are scaled by flow magnitude. Colormaps indicate the $X$ velocity component.}
  \end{center}
\end{figure}

\begin{figure}
  \begin{center}
    \includegraphics[width=.49\textwidth]{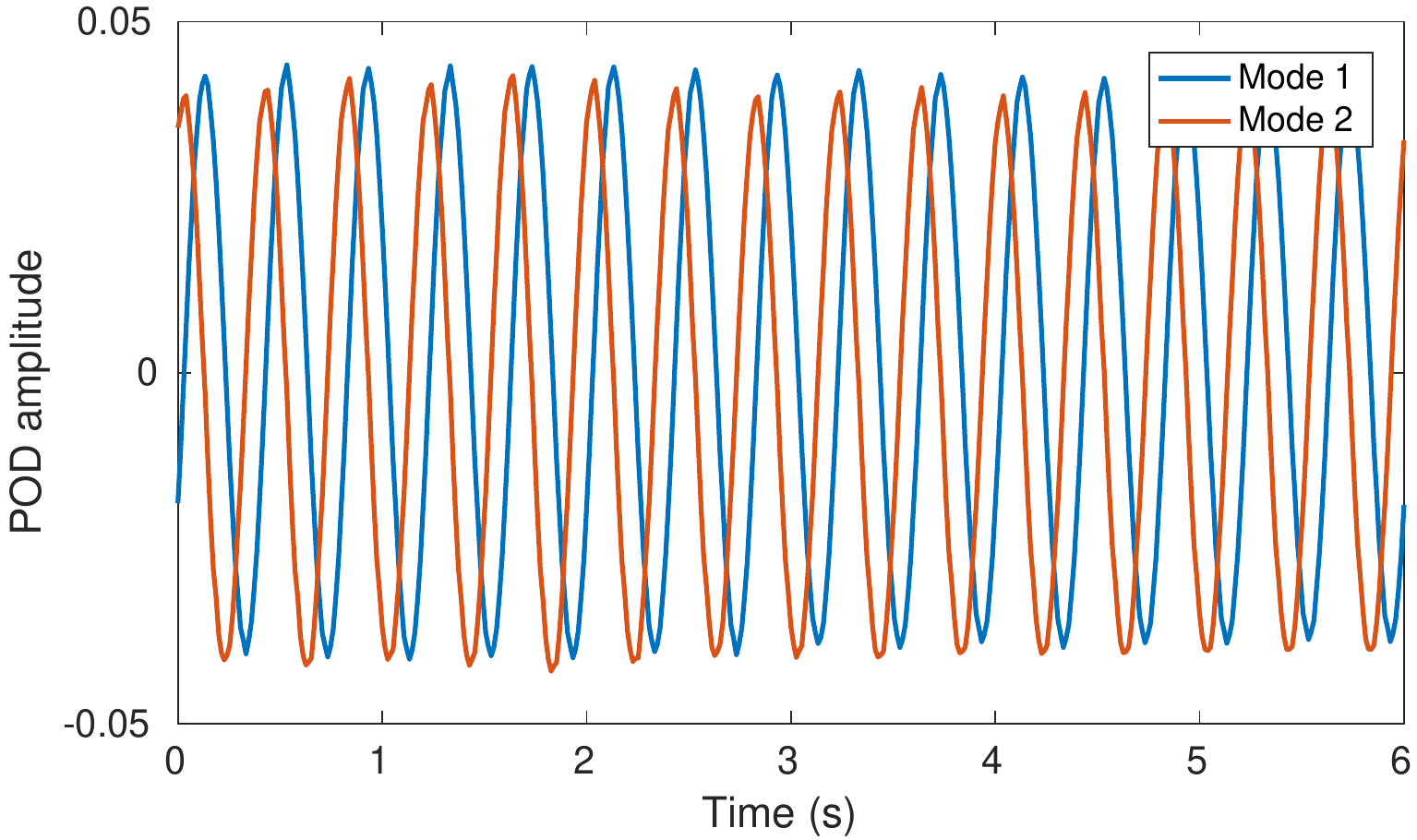}
    \includegraphics[width=.49\textwidth]{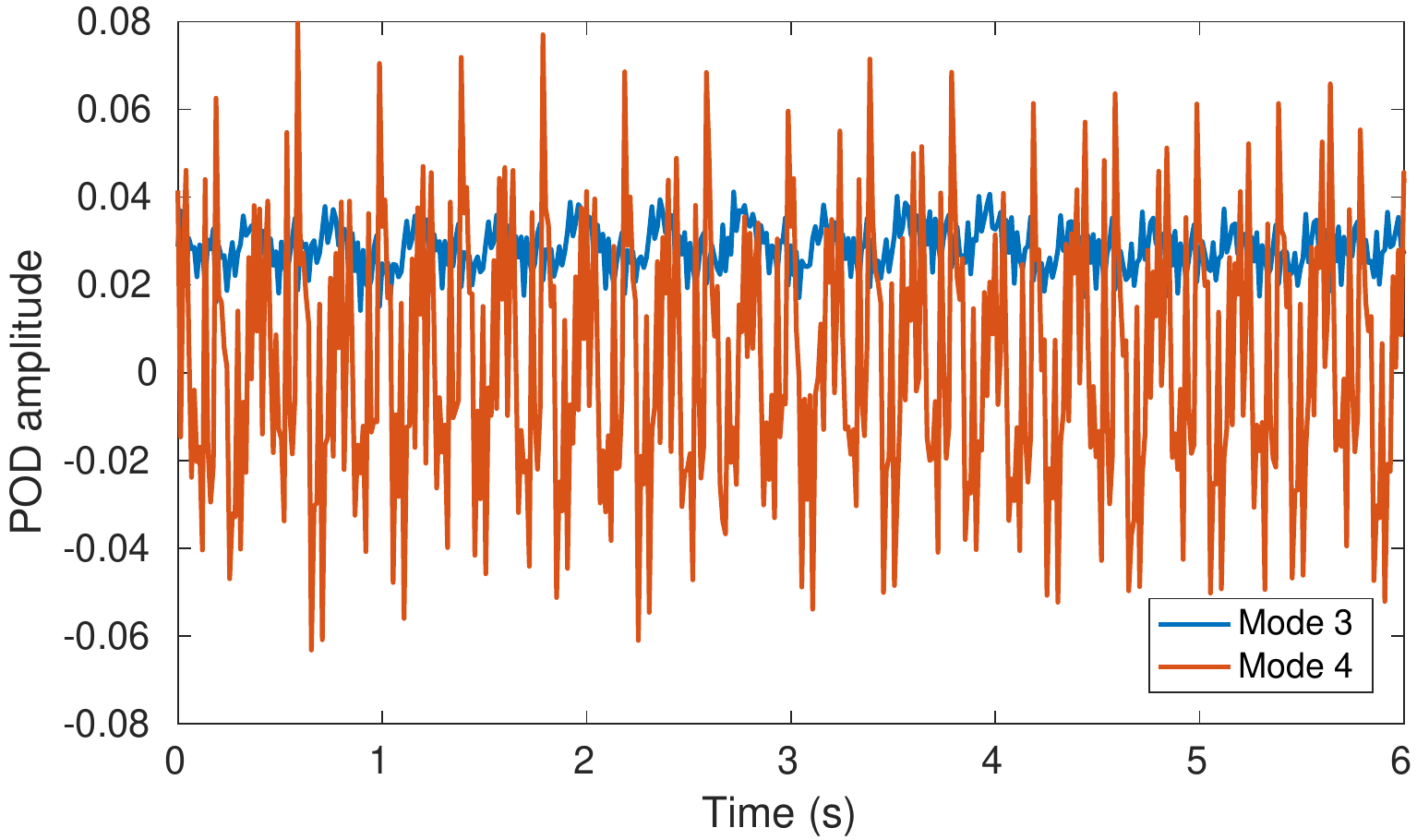}
    \caption{\label{case1_PODcoefficients} POD time coefficients in case 1, associated with mode 1 and 2 (left), and 3 and 4 (right).}
  \end{center}
\end{figure}

We first focus on the analysis of the POD modes obtained in case 1. The POD modes 1 and 2 obtained from the case 1 are presented in Fig. \ref{case1_PIV_continuous_mode12}. The POD modes 3 and 4, also in the case 1, are presented in Fig. \ref{case1_PIV_continuous_mode34}. The time coefficients of modes 1, 2, 3 and 4 in case 1 are presented in Fig. \ref{case1_PODcoefficients}. As visible on Fig. \ref{case1_PIV_continuous_mode12}, modes 1 and 2 are very similar. The plots showing the maps of the $X$ and $Y$ velocity components of mode 1 clearly demonstrate that those modes capture the orbital wave motion. Both modes 1 and 2 have nearly identical velocity magnitude fields, while being shifted spatially by a quarter of a wavelength when it regards the phase of the orbital motion they feature. This is confirmed by Fig. \ref{case1_PODcoefficients}, on which the time shift between the time coefficients for modes 1 and 2 is a quarter of a period.

This time-shift provides strong evidence that modes 1 and 2 represent the orthogonal basis for the wave orbital motion. This corresponds well with the fact that, in case 1, $87$ percent of the total energy is contained in modes 1 and 2. The time coefficients for modes 3 and 4, visible in Fig. \ref{case1_PODcoefficients}, are quite different from those for modes 1 and 2. The time coefficient for mode 3 does present some fluctuations with a 2.5~Hz frequency, but has a predominantly constant offset. Comparing this constant trend with the results presented in Fig. \ref{case1_PIV_continuous_mode34}, it appears that mode 3 describes a counter-flow under the grease ice layer. This is consistent with Fig. 19 of \citet{Martin1981}, who observed a counter-flowing grease ice flow that we could expect to trigger mean water currents under the ice. This can also be understood as a consequence of the packing of the ice by the wave-induced stress as the conservation of mass then requires a back-flow of water in the opposite direction. Mode 4 (Fig. \ref{case1_PIV_continuous_mode34}), in contrast, features no mean motion in the water domain, but a clear oscillatory motion at the interface with the grease ice. The time coefficient associated with this mode (Fig. \ref{case1_PODcoefficients}), though noisy, has a spectral peak at the wave frequency.

\begin{figure}
  \begin{center}
    \includegraphics[width=.49\textwidth]{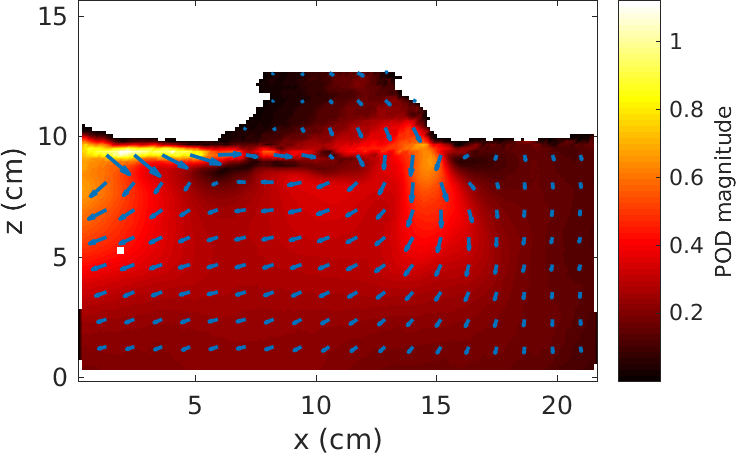}
    \includegraphics[width=.49\textwidth]{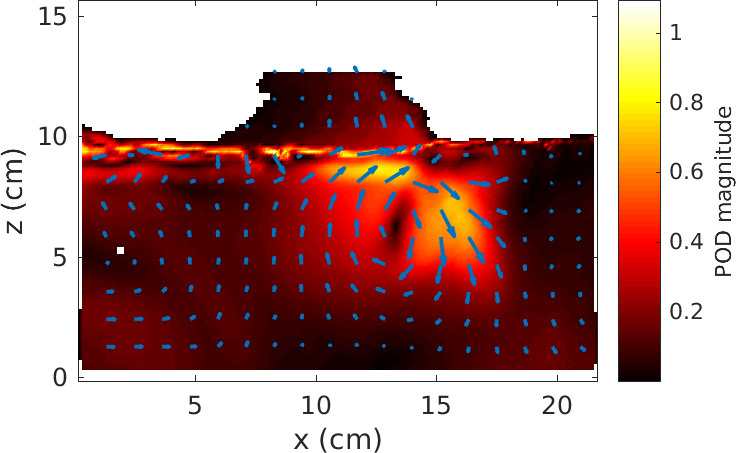}
    \includegraphics[width=.49\textwidth]{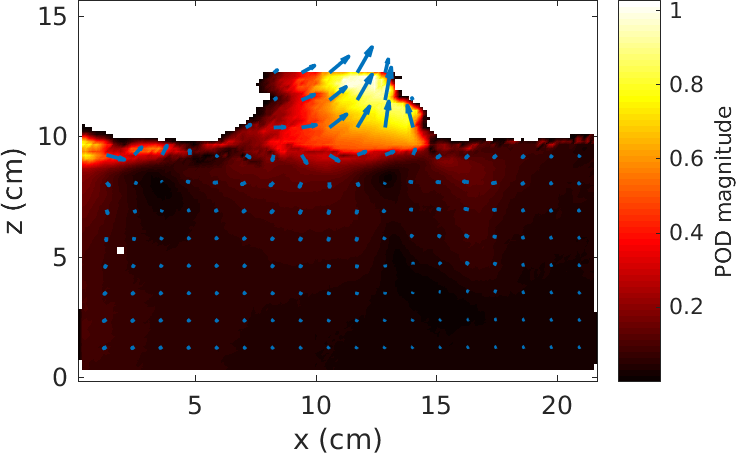}
    \includegraphics[width=.49\textwidth]{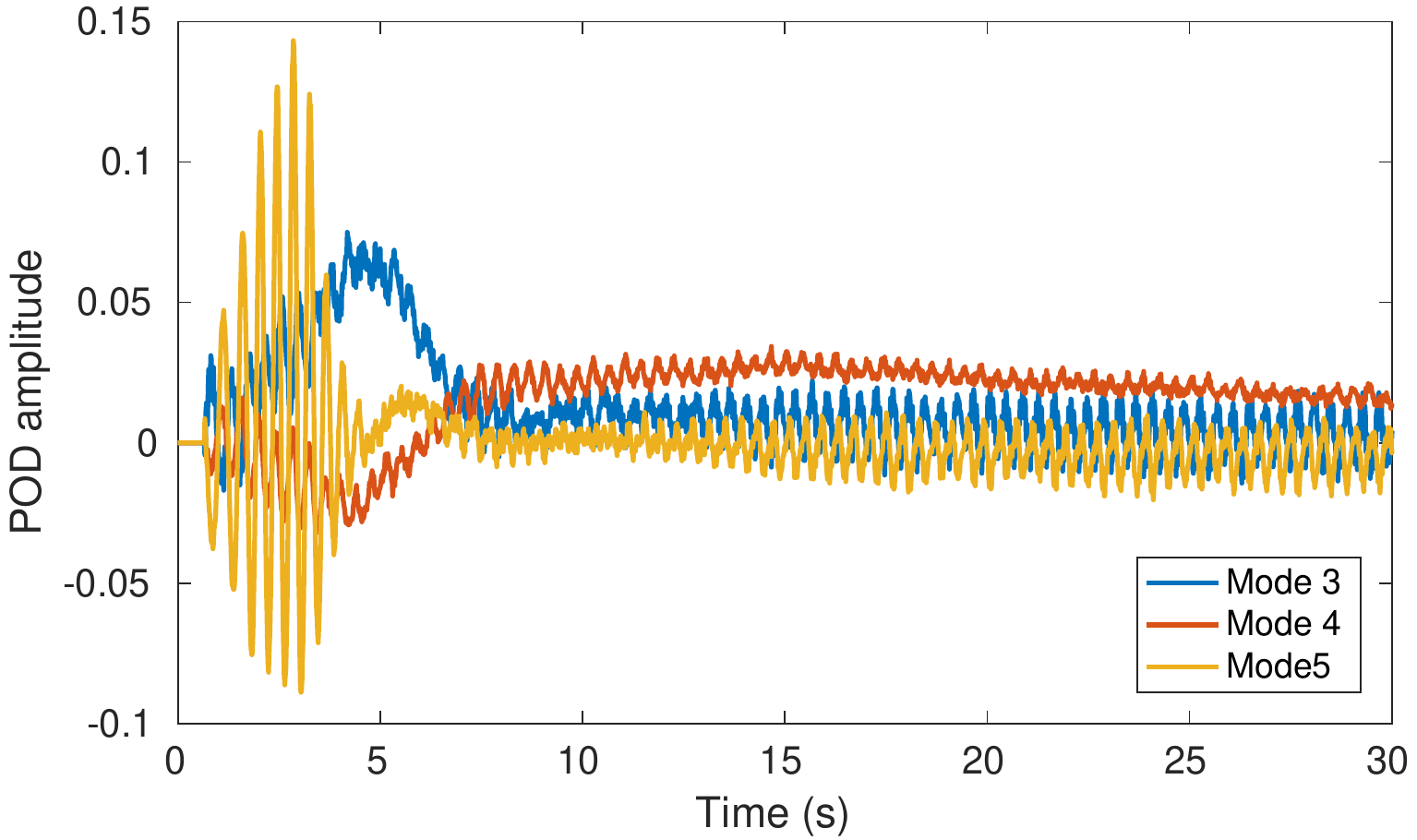}
    \caption{\label{case2_POD} POD modes 3, 4, 5 (first three plots, left to right, top to bottom) obtained in the case 2. Arrows indicate flow direction and are scaled by the velocity magnitude, the color map indicates velocity magnitude. Corresponding POD time coefficients (bottom right).}
  \end{center}
\end{figure}

POD modes are also computed in case 2. Modes 1 and 2 and the associated time coefficients are very similar to the results obtained in case 1, and are therefore not reproduced. Modes 3, 4 and 5, along with the corresponding time coefficients, are presented in Fig. \ref{case2_POD}. As visible on the POD time coefficients, a transient is observed at the beginning of the wave activity for about $7.5$ seconds, before a steady state is reached. Modes 3 and 5 are active in the beginning of the time series, and are essentially zero after $7.5$ seconds. By contrast, mode 4 gets active at around $7$ seconds, and slowly decays later on.

As was indicated previously, case 2 is chosen so as to investigate the transient effect induced by an initially inhomogeneous grease ice layer. This temporal evolution is  visible in the higher order POD modes presented in Fig. \ref{case2_POD}. The initial position of the grease ice is visible through the irregular shape of the interface at the top of the water. Initially, two packs of grease ice are present, with free water in between. As the waves develop, the grease ice pack on the left accelerates to the right under the influence of the waves, and the free water gap gets filled with grease ice. This leads to a collision between the two packs of grease ice creating a large vortex in the water. POD mode 3 corresponds to the fluid motion induced by the displacement of the left pack of grease ice. It is similar to the dipole-aspect fluid motion that would be obtained from the displacement of a solid rectangle, in potential flow theory. POD mode 4 is the vortex generated by the collision between the two grease ice packs. POD mode 5 is related to the jets generated before the time when the free water gap gets closed, due to the variations in the free water gap created by wave motion. The interpretation of those 3 modes is consistent with the temporal evolution of the respective POD modes.

\section{Discussion of effective viscosity and scaling of experiments}

The use of an effective water viscosity much higher than the kinematic viscosity of water in order to reproduce observations of wave damping is often discussed in the literature. While most authors justify it from considerations about water turbulence and the presence of eddies \citep{DeCarolis2002399}, the origin of those eddy structures, especially under grease-pancake ice where breaking waves are quickly attenuated, is problematic. The eddy structures could arise from wave-induced turbulence, but this topic is after many years still the object of much debate \citep{doi:10.1175/2009JPO4202.1, GRL:GRL22071, Beya2012}. At least two other mechanisms could participate in the generation of eddy structures, influencing wave damping by ice. Ice drifting relatively to the underlying water, for example under the influence of wind or waves, could create shear on large temporal and spatial scales and be the origin of eddy structures. Collisions between pancakes, small broken ice floes or small packs of grease ice could also inject energy in the superficial water layer. We document for the first time the existence of such eddy structures in a laboratory experiment, when the waves are started from a situation where some packs of grease ice are separated by open water regions. Unsurprisingly, collision between packs of grease ice leads to the creation of a jet when water is forced out of the gap, and a strong vortex is created which decays slowly with time. Such collision mechanisms could be important in the field for injecting energy in the water under the ice, and therefore increasing the level of eddy viscosity and viscous dissipation.

However, more work is necessary to confirm this mechanism. Indeed, scaling between real sea conditions and wave tank experiments is problematic, and the generality of the experiment presented here would need to be further validated in experiments of larger scale.

More specifically, the very first imperative of laboratory experiments on waves is to be in the right water depth regime. Due to the very limited depth of most wave tanks, this puts a sharp requirement on the minimum frequency that can be used for tests in the laboratory. We consider as a first approximation in all the discussion that follows that the deep water dispersion relation is enforced, so that $\omega^2 = gk$, and that deep-water waves are the most common situation for the ocean, which implies that $kH > 1$, i.e. $\omega > \omega_{min} = \sqrt{g/H}$. In addition, the steepness of the waves must be kept moderate to limit nonlinear effects, typically $ \epsilon = k a \approx 0.1$, i.e. $a < 0.1 g / \omega^2$, where $k$ is the wave vector and $a$ the wave amplitude. If the amplitude-based Reynolds number defined in \cite{GRL:GRL22071}, $Re_a = a^2 \omega / \nu_{water}$, is the right non-dimensional parameter for describing turbulence under waves and large eddy structures under the ice, then in a laboratory experiment, with a wave-tank of depth H:

\begin{equation}
  Re_a = \frac{a^2 \omega}{\nu_{water}} < \frac{10^{-2} g^2}{\omega^3 \nu_{water}} < \frac{10^{-2} g^{1/2} H^{3/2}}{\nu_{water}}.
\end{equation}

This implies that, even in the case of a wave-tank several meters deep, the maximum Reynolds number that can be achieved in the laboratory for deep water waves of reasonable steepness is much lower than any typical swell on deep water. More specifically, the maximum Reynolds number obtained in a wave tank scales as the water depth to the power $3/2$. As a consequence, studying phenomena related to water wave turbulence and wave-induced eddy viscosity is challenging in the laboratory, as it is impossible to investigate whether a higher Reynolds number could lead to different physics. However, while this scaling issue is certainly problematic in the study of, for example, wave turbulence, one could expect that it is less of a problem in the present case. Indeed, the dynamics presented here are not relying on the existence of a general instability in the flow, but rather on largely inviscid dynamics, with the effect of viscosity being limited to thin boundary layers that are difficult to observe as they are predicted to be on the order of $\delta = \sqrt{2 \nu / \omega} \approx 5 \times 10^{-4}$~m, using an approximate value for water viscosity $\nu = 1.7 \times 10^{-6}$~m$^2$/s and a frequency $f = \omega / (2 \pi) = 2.0$~Hz. For example, the eddy we observed is due to the existence of a water jet created by the reduction of the free volume between the grease ice packs, rather than (directly) viscous effects. In addition, the dynamics analyzed presently are relevant for the grease ice part of the Marginal Ice Zone, where low frequency waves propagate mostly unaffected and high frequency waves, such as those investigated here, are the ones interacting most with the ice.

\section{Conclusions}

A set of experiments about wave propagation in grease ice is presented. For the first time, the data reported contain both traditional single point measurements obtained with ultrasonic gauges, and measurements of the water kinematics using Particle Image Velocimetry. In addition, the frequency range investigated is much larger than what was reported in previous experiments. Collecting such data is necessary to help progress towards the development of models describing propagation of waves in the Marginal Ice Zone.

In good agreement with \citet{Martin1981}, we observe that the grease-ice layer gets thicker in the direction of wave propagation, which can be understood from the effect of the gradient in the wave-induced stress. In addition, we are able to measure with PIV the recirculation flow under the grease ice, imposed by mass conservation as a result of the grease ice layer thickening along the direction of wave propagation and also visually observed by \citet{Martin1981}. Unfortunately, the distribution of grease ice thickness during the experiments, and its variation with time and position, was not measured directly. Therefore no detailed comparison can be drawn between the grease ice thickness predictions of the mass loading and damping models, and reality. This indicates the necessity to perform detailed time measurements of grease ice thickness in subsequent experiments. When optical access is granted, this could be performed by recording calibrated side images of the grease ice layer.

By contrast with the findings by \citet{NewyearLab1}, we observe an increase in the wavenumber (reduction in the wavelength) in a way similar to what is described by the mass-loading model for intermediate frequencies. This phenomenon gets more visible in the higher frequency range we investigate than in the lower frequency range used by \citet{NewyearLab1}, which may explain why this observation was not reported previously. In addition, the wavenumbers observed depend upon the position along the wave-tank as a consequence of the grease ice layer thickness gradient previously described. For the highest part of the frequency range considered, when the wavelength is no longer much larger than the grease ice layer thickness, the wavenumber reaches a plateau. This corresponds to the fact that the grease ice can no longer be considered as a thin layer on top of the water, and therefore the hypothesis used in the mass loading model is no longer valid.

Wave attenuation is maybe the most important parameter to predict for real world applications. PIV and ultrasonic gauges measurements confirm the existence of an exponential wave attenuation when waves propagate in the ice. Interestingly, we observe a transition in the curve for the attenuation coefficient at around 2.2 Hz, relative to the simple model by \citet{WeberArticle}. Such feature had not been reported as clearly in the literature before, and we can observe it as a consequence of the wide frequency range investigated. This transition is successfully captured by the model of \citet{SutherlandDissipation}, adding credibility to the physical explanation it relies on.

Finally, PIV measurements reveal the existence of an active eddy under the ice when discontinuous patches of grease ice collide with each other under the influence of waves. Although a simple and unsurprising phenomenon, this is the first time to the authors knowledge that an eddy generation mechanism under sea ice has been documented in a laboratory experiment. The underlying collision mechanism may be important in real world conditions in the case of tightly packed ice.

\section{Acknowledgements}

We want to thank laboratory engineer Olav Gundersen for his help and precious advices when designing the wave maker. We are grateful to UNIS faculty and staff, for their help and hospitality during the experimental campaign in Svalbard. The help of Jostein Kolaas, whose Matlab scripts are often an inspiration for other members of the laboratory when processing Digiflow data, is acknowledged. This study was funded by the Norwegian Research Council under the PETROMAKS2 scheme [project WOICE, Grant Number 233901, and project DOFI, Grant number 280625].

\bibliography{BibliographyBib}

\begin{thebibliography}{42}
\expandafter\ifx\csname natexlab\endcsname\relax\def\natexlab#1{#1}\fi
\def\au#1{#1} \def\ed#1{#1} \def\yr#1{#1}\def\at#1{#1}\def\jt#1{\textit{#1}}
  \def\bt#1{#1}\def\bvol#1{\textbf{#1}} \def\vol#1{#1} \def\pg#1{#1}
  \def\publ#1{#1}\def\arxiv#1{#1}\def\org#1{#1}\def\st#1{\textit{#1}}

\bibitem[Babanin(2006)]{GRL:GRL22071}
{\sc \au{Babanin, A.~V.}} \yr{2006}  \at{On a wave-induced turbulence and a
  wave-mixed upper ocean layer}.  \jt{Geophysical Research Letters}
  \bvol{33}~(20),  \pg{n/a--n/a}, l20605.

\bibitem[Babanin \& Haus(2009)]{doi:10.1175/2009JPO4202.1}
{\sc \au{Babanin, Alexander~V.} \& \au{Haus, Brian~K.}} \yr{2009}  \at{On the
  existence of water turbulence induced by nonbreaking surface waves}.
  \jt{Journal of Physical Oceanography}  \bvol{39}~(10),  \pg{2675--2679},
  \arxiv{arXiv: http://dx.doi.org/10.1175/2009JPO4202.1}.

\bibitem[Berkooz {\em et~al.\/}(1993)Berkooz, Holmes \&
  Lumley]{berkooz1993proper}
{\sc \au{Berkooz, Gal}, \au{Holmes, Philip} \& \au{Lumley, John~L}} \yr{1993}
  \at{The proper orthogonal decomposition in the analysis of turbulent flows}.
  \jt{Annual review of fluid mechanics}  \bvol{25}~(1),  \pg{539--575}.

\bibitem[Bey{\'a} {\em et~al.\/}(2012)Bey{\'a}, Peirson \& Banner]{Beya2012}
{\sc \au{Bey{\'a}, J.~F.}, \au{Peirson, W.~L.} \& \au{Banner, M.~L.}} \yr{2012}
   \at{Turbulence beneath finite amplitude water waves}.  \jt{Experiments in
  Fluids}  \bvol{52}~(5),  \pg{1319--1330}.

\bibitem[Brostr\"om \& Christensen(2008)]{KaiReport}
{\sc \au{Brostr\"om, G\"oran} \& \au{Christensen, Kai}} \yr{2008}  \bt{Waves in
  sea ice}. {\em Tech. Rep.\/}~5.  \org{Norwegian Meteorological Institute}.

\bibitem[Collins {\em et~al.\/}(2017)Collins, Rogers \&
  Lund]{Collins2017dispersioninice}
{\sc \au{Collins, Clarence~Olin}, \au{Rogers, William~Erick} \& \au{Lund,
  Bj{\"o}rn}} \yr{2017}  \at{An investigation into the dispersion of ocean
  surface waves in sea ice}.  \jt{Ocean Dynamics}  \bvol{67}~(2),
  \pg{263--280}.

\bibitem[Collins {\em et~al.\/}(2015)Collins, Rogers, Marchenko \&
  Babanin]{GRL:GRL52708}
{\sc \au{Collins, Clarence~O.}, \au{Rogers, W.~Erick}, \au{Marchenko, Aleksey}
  \& \au{Babanin, Alexander~V.}} \yr{2015}  \at{In situ measurements of an
  energetic wave event in the arctic marginal ice zone}.  \jt{Geophysical
  Research Letters}  \bvol{42}~(6),  \pg{1863--1870}, 2015GL063063.

\bibitem[Dalziel(2012)]{Digiflow}
{\sc \au{Dalziel, Stuart}} \yr{2012} {\em DigiFlow user guide\/}, {3.4} edn.
  Dalziel research partners,
  \url{http://www.dalzielresearch.com/digiflow/digiflow.pdf}.

\bibitem[De~Carolis \& Desiderio(2002)]{DeCarolis2002399}
{\sc \au{De~Carolis, Giacomo} \& \au{Desiderio, Daniela}} \yr{2002}
  \at{Dispersion and attenuation of gravity waves in ice: a two-layer viscous
  fluid model with experimental data validation}.  \jt{Physics Letters A}
  \bvol{305}~(6),  \pg{399 -- 412}.

\bibitem[Doble {\em et~al.\/}(2015)Doble, De~Carolis, Meylan, Bidlot \&
  Wadhams]{GRL:GRL53001}
{\sc \au{Doble, Martin~J.}, \au{De~Carolis, Giacomo}, \au{Meylan, Michael~H.},
  \au{Bidlot, Jean-Raymond} \& \au{Wadhams, Peter}} \yr{2015}  \at{Relating
  wave attenuation to pancake ice thickness, using field measurements and model
  results}.  \jt{Geophysical Research Letters}  \bvol{42}~(11),
  \pg{4473--4481}, 2015GL063628.

\bibitem[Gaster(1962)]{gaster_1962}
{\sc \au{Gaster, M.}} \yr{1962}  \at{A note on the relation between
  temporally-increasing and spatially-increasing disturbances in hydrodynamic
  stability}.  \jt{Journal of Fluid Mechanics}  \bvol{14}~(2),  \pg{222–224}.

\bibitem[Joshua(2014)]{OSLab}
{\sc \au{Joshua, M.~P.}} \yr{2014} {\em Open-Source Lab\/}.  \publ{Elsevier}.

\bibitem[{Keller}(1998)]{TwoLayersModel}
{\sc \au{{Keller}, J.~B.}} \yr{1998}  \at{Gravity waves on ice-covered water}.
  \jt{Journal of Geophysical Research}  \bvol{103},  \pg{7663--7669}.

\bibitem[Kerschen {\em et~al.\/}(2005)Kerschen, Golinval, Vakakis \&
  Bergman]{Kerschen2005}
{\sc \au{Kerschen, Gaetan}, \au{Golinval, Jean-claude}, \au{Vakakis,
  Alexander~F.} \& \au{Bergman, Lawrence~A.}} \yr{2005}  \at{The method of
  proper orthogonal decomposition for dynamical characterization and order
  reduction of mechanical systems: An overview}.  \jt{Nonlinear Dynamics}
  \bvol{41}~(1),  \pg{147--169}.

\bibitem[Lamb(1932)]{lamb1932hydrodynamics}
{\sc \au{Lamb, H.}} \yr{1932} {\em Hydrodynamics\/}.  \publ{Cambridge
  University Press}.

\bibitem[Li {\em et~al.\/}(2017)Li, Kohout, Doble, Wadhams, Guan \&
  Shen]{li2017rollover}
{\sc \au{Li, Jingkai}, \au{Kohout, Alison~L}, \au{Doble, Martin~J},
  \au{Wadhams, Peter}, \au{Guan, Changlong} \& \au{Shen, Hayley~H}} \yr{2017}
  \at{Rollover of apparent wave attenuation in ice covered seas}.  \jt{Journal
  of Geophysical Research: Oceans}  \bvol{122}~(11),  \pg{8557--8566}.

\bibitem[Marchenko {\em et~al.\/}(2017)Marchenko, Rabault, Sutherland, Collins,
  Wadhams \& Chumakov]{marchenko2017field}
{\sc \au{Marchenko, A.}, \au{Rabault, J.}, \au{Sutherland, G.}, \au{Collins, C.
  O.~III}, \au{Wadhams, P} \& \au{Chumakov, M.}} \yr{2017} Field observations
  and preliminary investigations of a wave event in solid drift ice in the
  barents sea.  \bt{In {\em 24th International Conference on Port and Ocean
  Engineering under Arctic Conditions\/}}.

\bibitem[Martin \& Kauffman(1981)]{Martin1981}
{\sc \au{Martin, S.} \& \au{Kauffman, P.}} \yr{1981}  \at{A field and
  laboratory study of wave damping by grease ice}.  \jt{Journal of Glaciology}
  \bvol{27},  \pg{283--313}.

\bibitem[Meylan {\em et~al.\/}(2018)Meylan, Bennetts, Mosig, Rogers, Doble \&
  Peter]{meylan2018dispersion}
{\sc \au{Meylan, MH}, \au{Bennetts, LG}, \au{Mosig, JEM}, \au{Rogers, WE},
  \au{Doble, MJ} \& \au{Peter, MA}} \yr{2018}  \at{Dispersion relations, power
  laws, and energy loss for waves in the marginal ice zone}.  \jt{Journal of
  Geophysical Research: Oceans} .

\bibitem[Mosig {\em et~al.\/}(2015)Mosig, Montiel \& Squire]{JGRC:JGRC21350}
{\sc \au{Mosig, Johannes E.~M.}, \au{Montiel, Fabien} \& \au{Squire,
  Vernon~A.}} \yr{2015}  \at{Comparison of viscoelastic-type models for ocean
  wave attenuation in ice-covered seas}.  \jt{Journal of Geophysical Research:
  Oceans}  \bvol{120}~(9),  \pg{6072--6090}.

\bibitem[Newyear \& Martin(1997)]{NewyearLab1}
{\sc \au{Newyear, Karl} \& \au{Martin, Seelye}} \yr{1997}  \at{A comparison of
  theory and laboratory measurements of wave propagation and attenuation in
  grease ice}.  \jt{Journal of Geophysical Research: Oceans}  \bvol{102}~(C11),
   \pg{25091--25099}.

\bibitem[Newyear \& Martin(1999)]{NewyearLab2}
{\sc \au{Newyear, Karl} \& \au{Martin, Seelye}} \yr{1999}  \at{Comparison of
  laboratory data with a viscous two-layer model of wave propagation in grease
  ice}.  \jt{Journal of Geophysical Research: Oceans}  \bvol{104}~(C4),
  \pg{7837--7840}.

\bibitem[Pfirman {\em et~al.\/}(1995)Pfirman, Eicken, Bauch \&
  Weeks]{Pfirman1995129}
{\sc \au{Pfirman, S.L.}, \au{Eicken, H.}, \au{Bauch, D.} \& \au{Weeks, W.F.}}
  \yr{1995}  \at{The potential transport of pollutants by arctic sea ice}.
  \jt{Science of The Total Environment}  \bvol{159}~(2–3),  \pg{129 -- 146}.

\bibitem[Rabault {\em et~al.\/}(2016)Rabault, Halsne, Sutherland \&
  Jensen]{rabault2016ptv}
{\sc \au{Rabault, J}, \au{Halsne, T}, \au{Sutherland, G} \& \au{Jensen, A}}
  \yr{2016} {PTV} investigation of the mean drift currents under water waves.
  \bt{In {\em Proceedings of the 18th Int. Lisb. Symp.\/}}.

\bibitem[Rabault {\em et~al.\/}(2017)Rabault, Sutherland, Gundersen \&
  Jensen]{RabaultSutherlandGlaciology}
{\sc \au{Rabault, Jean}, \au{Sutherland, Graig}, \au{Gundersen, Olav} \&
  \au{Jensen, Atle}} \yr{2017}  \at{Measurements of wave damping by a grease
  ice slick in svalbard using off-the-shelf sensors and open-source
  electronics}.  \jt{Journal of Glaciology}  \pg{p. 1–10}.

\bibitem[Rao(1973)]{rao73}
{\sc \au{Rao, C.~Radhakrishna}} \yr{1973} {\em Linear Statistical Inference and
  its Applications\/}.  \publ{Wiley}.

\bibitem[{Smedsrud}(2011)]{2011AnGla5277S}
{\sc \au{{Smedsrud}, L.~H.}} \yr{2011}  \at{{Grease-ice thickness
  parameterization}}.  \jt{Annals of Glaciology}  \bvol{52},  \pg{77--82}.

\bibitem[Smedsrud \& Skogseth(2006)]{Smedsrud2006171}
{\sc \au{Smedsrud, Lars~H.} \& \au{Skogseth, Ragnheid}} \yr{2006}  \at{Field
  measurements of arctic grease ice properties and processes}.  \jt{Cold
  Regions Science and Technology}  \bvol{44}~(3),  \pg{171 -- 183}.

\bibitem[Squire {\em et~al.\/}(1995)Squire, Dugan, Wadhams, Rottier \&
  K.]{SquireOOWASI}
{\sc \au{Squire, V.A.}, \au{Dugan, J.~P.}, \au{Wadhams, P.}, \au{Rottier,
  P.~J.} \& \au{K., A.}} \yr{1995}  \at{Of ocean waves and sea-ice}.
  \jt{Annual Review of Fluid Mechanics}  \bvol{27},  \pg{115 -- 168}.

\bibitem[Squire \& Montiel(2016)]{squire2016evolution}
{\sc \au{Squire, Vernon~A} \& \au{Montiel, Fabien}} \yr{2016}  \at{Evolution of
  directional wave spectra in the marginal ice zone: A new model tested with
  legacy data}.  \jt{Journal of Physical Oceanography}  \bvol{46}~(10),
  \pg{3121--3137}.

\bibitem[Sutherland {\em et~al.\/}(2018)Sutherland, Christensen, Rabault \&
  Jensen]{SutherlandDissipation}
{\sc \au{Sutherland, Graig}, \au{Christensen, Kai~H.}, \au{Rabault, Jean} \&
  \au{Jensen, Atle}} \yr{2018}  \at{A new look at wave dissipation in the
  marginal ice zone}.  \jt{Under review, available at
  https://arxiv.org/pdf/1805.01134.pdf} .

\bibitem[Sutherland {\em et~al.\/}(2017)Sutherland, Halsne, Rabault \&
  Jensen]{Sutherland201788}
{\sc \au{Sutherland, Graig}, \au{Halsne, Trygve}, \au{Rabault, Jean} \&
  \au{Jensen, Atle}} \yr{2017}  \at{The attenuation of monochromatic surface
  waves due to the presence of an inextensible cover}.  \jt{Wave Motion}
  \bvol{68},  \pg{88 -- 96}.

\bibitem[Sutherland \& Rabault(2016)]{JGRC:JGRC21649}
{\sc \au{Sutherland, Graig} \& \au{Rabault, Jean}} \yr{2016}  \at{Observations
  of wave dispersion and attenuation in landfast ice}.  \jt{Journal of
  Geophysical Research: Oceans}  \bvol{121}~(3),  \pg{1984--1997}.

\bibitem[Thomson \& Rogers(2014)]{GRL:GRL51656}
{\sc \au{Thomson, Jim} \& \au{Rogers, W.~Erick}} \yr{2014}  \at{Swell and sea
  in the emerging arctic ocean}.  \jt{Geophysical Research Letters}
  \bvol{41}~(9),  \pg{3136--3140}.

\bibitem[Van~Dorn(1966)]{van1966boundary}
{\sc \au{Van~Dorn, WG}} \yr{1966}  \at{Boundary dissipation of oscillatory
  waves}.  \jt{Journal of Fluid Mechanics}  \bvol{24}~(04),  \pg{769--779}.

\bibitem[Wadhams(1973)]{wadhams1973attenuation}
{\sc \au{Wadhams, Peter}} \yr{1973}  \at{Attenuation of swell by sea ice}.
  \jt{Journal of Geophysical Research}  \bvol{78}~(18),  \pg{3552--3563}.

\bibitem[Wadhams \& Doble(2009)]{Wadhams200998}
{\sc \au{Wadhams, Peter} \& \au{Doble, Martin~J.}} \yr{2009}  \at{Sea ice
  thickness measurement using episodic infragravity waves from distant storms}.
   \jt{Cold Regions Science and Technology}  \bvol{56}~(2–3),  \pg{98 --
  101}.

\bibitem[Wadhams {\em et~al.\/}(1988)Wadhams, Squire, Goodman, Cowan \&
  Moore]{JGRC:JGRC4212}
{\sc \au{Wadhams, Peter}, \au{Squire, Vernon~A.}, \au{Goodman, Dougal~J.},
  \au{Cowan, Andrew~M.} \& \au{Moore, Stuart~C.}} \yr{1988}  \at{The
  attenuation rates of ocean waves in the marginal ice zone}.  \jt{Journal of
  Geophysical Research: Oceans}  \bvol{93}~(C6),  \pg{6799--6818}.

\bibitem[Wang \& Shen(2010{\natexlab{{\em a\/}}})]{Wang201090}
{\sc \au{Wang, Ruixue} \& \au{Shen, Hayley~H.}} \yr{2010{\natexlab{{\em a\/}}}}
   \at{Experimental study on surface wave propagating through a
  grease–pancake ice mixture}.  \jt{Cold Regions Science and Technology}
  \bvol{61}~(2 - 3),  \pg{90 -- 96}.

\bibitem[Wang \& Shen(2010{\natexlab{{\em b\/}}})]{JGRC:JGRC11467}
{\sc \au{Wang, Ruixue} \& \au{Shen, Hayley~H.}} \yr{2010{\natexlab{{\em b\/}}}}
   \at{Gravity waves propagating into an ice-covered ocean: A viscoelastic
  model}.  \jt{Journal of Geophysical Research: Oceans}  \bvol{115}~(C6).

\bibitem[Weber(1987)]{WeberArticle}
{\sc \au{Weber, Jan~Erik}} \yr{1987}  \at{Wave attenuation and wave drift in
  the marginal ice zone}.  \jt{Journal of Physical Oceanography}
  \bvol{17}~(12),  \pg{2351--2361},  \arxiv{arXiv:
  http://dx.doi.org/10.1175/1520-0485(1987)017<2351:WAAWDI>2.0.CO;2}.

\bibitem[Zhao \& Shen(2015)]{Zhao201571}
{\sc \au{Zhao, Xin} \& \au{Shen, Hayley~H.}} \yr{2015}  \at{Wave propagation in
  frazil/pancake, pancake, and fragmented ice covers}.  \jt{Cold Regions
  Science and Technology}  \bvol{113},  \pg{71 -- 80}.

\end{thebibliography}
\bibliographystyle{jfm}

\section{Appendix A: technical details of the wave paddle and gauges setup}

Both the actuation system used to drive the wave paddle and the logging system used to record the ultrasonic gauges signal are released as open source material. The specificity of open source software and electronics is that their source code, internal designs and interfaces are made available through a license that provides the right to study, modify and distribute the product \citep{OSLab}. This has many valuable implications for the scientific community. In particular, sharing all the details of the design of an instrument or experiment setup can make it easier to reproduce experiments, by substantially reducing the cost and time necessary to build an exact copy of the instrument initially used. In addition, this makes it easier to build upon a common platform, therefore encouraging modularity and reuse of previous designs rather than fragmented in-house development, which very likely is redundant between research groups and private suppliers, leading to unnecessary costs.

Both the paddle control system and the gauges logging system are based on Arduino boards. The paddle control system relies on an Arduino Due that works at 3.3V and features a 12 bits Analog to Digital Converter (ADC), while the gauges logging system relies on an Arduino Mega working at 5V and using a 10 bits ADC. Previously, our group used more expansive acquisition boards from Texas Instruments for performing this logging, but the gauges rather than the acquisition boards are the main source of noise as explained in the next paragraph and we could not see any difference in signal quality between the TI acquisition boards and the 20 to 30 times less expansive Arduino Mega on this specific task. The Arduino Due core program is modified so that extended serial buffers can be used. As a consequence, we can transmit from the computer the necessary position instruction buffers that ensure a control frequency of 500Hz, despite the small latency of the computer Operating System. The Arduino Due itself runs a Proportional Integral Differential (PID) control loop at a frequency of 2kHz. The control loop is designed so that the paddle position, which is measured by a LVDT sensor, follows to the values transmitted in the buffers.

Regarding wave elevation measurements, the amplitude of the signals generated by each ultrasonic gauge (going from 0 to 10V) is first divided by 2 using a voltage divider, before being read by the Arduino Mega. The reading frequency is set to 100Hz. The data generated by the Arduino Mega is then transmitted to the logging computer by USB. Since the resolution of the gauges is $0.5$ mm, while the sensing range is between $30$ and $300$ mm, i.e. $270$mm wide, the 10 bits resolution of the Arduino Mega ADC is better than the resolution of the gauges. All gauges are calibrated before the experiments are performed, by recording the data provided by the Arduino Mega at 5 different, fixed water heights and using those values as reference points for a linear least squares fit.

Both the paddle actuation and the ultrasonic gauges measurements are controlled as two separated threads from the same program. This lets us fully automate the experiments, and as a consequence no human input is necessary to perform a set of measurements. This is the reason why we are able to provide a dataset featuring more points and covering a wider frequency range than previous study: experiments were run automatically, and therefore no more work is required to generate additional measurements once the system is set. All the details of the actuation and logging systems are provided on the Github of the corresponding author (https://github.com/jerabaul29/PaddleAndUltrasonicGauges).

\section{Appendix B: wavenumber and damping coefficient data}

This Appendix contains the results obtained from the probes data following the methodology detailed in section 2. These data are used to generate Figs. \ref{wavelength} and \ref{fig_damping}. Some attenuation data at low frequency are of bad quality due to too low attenuation on the distance considered, and are therefore not reported (N/A) neither used in Fig. \ref{fig_damping}.

\begin{longtable}{|c|c|c|c|c|c|c|c|}
  \hline
  f (Hz) & a (mm) & $\alpha$ ($m^{-1}$) & $\Delta \alpha$ 95\% ($m^{-1}$) & $k_0$ ($m^{-1}$) & $k_{12}$ ($m^{-1}$) & $k_{23}$ ($m^{-1}$) & $k_{13}$ ($m^{-1}$) \\
  \hline\hline
  1.5 & 4.0 & N/A & N/A & 09.25 & 08.19 & 12.47 & 15.79 \\
1.5 & 4.4 & N/A & N/A & 09.25 & 07.67 & 12.24 & 15.78 \\
1.5 & 4.1 & N/A & N/A & 09.25 & 07.48 & 11.87 & 15.27 \\
1.5 & 4.6 & N/A & N/A & 09.25 & 05.68 & 11.20 & 15.47 \\
1.5 & 5.1 & N/A & N/A & 09.25 & 07.27 & 11.85 & 15.40 \\
1.5 & 5.7 & N/A & N/A & 09.25 & 07.18 & 12.11 & 15.94 \\
1.6 & 5.5 & N/A & N/A & 10.43 & 10.44 & 15.39 & 19.23 \\
1.6 & 6.1 & N/A & N/A & 10.43 & 10.70 & 15.00 & 18.32 \\
1.6 & 6.7 & N/A & N/A & 10.43 & 11.46 & 15.37 & 18.39 \\
1.6 & 7.1 & N/A & N/A & 10.43 & 11.03 & 15.16 & 18.37 \\
1.6 & 7.8 & N/A & N/A & 10.43 & 11.60 & 15.40 & 18.34 \\
1.6 & 8.3 & N/A & N/A & 10.43 & 10.98 & 14.68 & 17.55 \\
1.7 & 5.3 & 0.21 & 0.19 & 11.71 & 11.35 & 17.40 & 22.10 \\
1.7 & 5.9 & 0.22 & 0.19 & 11.71 & 11.49 & 17.45 & 22.08 \\
1.7 & 6.6 & 0.25 & 0.22 & 11.71 & 11.51 & 16.91 & 21.10 \\
1.7 & 7.1 & 0.24 & 0.21 & 11.71 & 10.41 & 16.97 & 22.06 \\
1.7 & 7.8 & 0.23 & 0.19 & 11.71 & 11.07 & 17.22 & 22.00 \\
1.7 & 8.4 & 0.24 & 0.20 & 11.71 & 11.66 & 17.53 & 22.08 \\
1.8 & 5.5 & 0.42 & 0.35 & 13.08 & 12.32 & 20.70 & 27.19 \\
1.8 & 6.1 & 0.44 & 0.35 & 13.08 & 13.17 & 21.29 & 27.58 \\
1.8 & 6.6 & 0.39 & 0.34 & 13.08 & 11.82 & 20.70 & 27.59 \\
1.8 & 7.2 & 0.40 & 0.32 & 13.08 & 12.54 & 20.59 & 26.84 \\
1.8 & 7.9 & 0.40 & 0.33 & 13.08 & 13.30 & 21.01 & 26.98 \\
1.8 & 8.5 & 0.34 & 0.30 & 13.08 & 12.62 & 20.81 & 27.16 \\
1.9 & 6.0 & 0.71 & 0.61 & 14.55 & 14.70 & 25.33 & 33.57 \\
1.9 & 6.7 & 0.70 & 0.57 & 14.55 & 15.36 & 25.54 & 33.44 \\
1.9 & 7.4 & 0.73 & 0.56 & 14.55 & 13.95 & 24.99 & 33.55 \\
1.9 & 7.9 & 0.68 & 0.58 & 14.55 & 14.81 & 25.35 & 33.52 \\
1.9 & 8.6 & 0.69 & 0.56 & 14.55 & 14.03 & 25.27 & 33.99 \\
1.9 & 9.4 & 0.68 & 0.57 & 14.55 & 12.52 & 24.36 & 33.55 \\
2.0 & 6.3 & 0.71 & 0.55 & 16.11 & 15.59 & 28.57 & 38.64 \\
2.0 & 6.9 & 0.74 & 0.51 & 16.11 & 16.17 & 28.87 & 38.73 \\
2.0 & 7.6 & 0.75 & 0.53 & 16.11 & 15.53 & 28.71 & 38.93 \\
2.0 & 8.2 & 0.74 & 0.54 & 16.11 & 15.40 & 28.76 & 39.12 \\
2.0 & 8.9 & 0.83 & 0.56 & 16.11 & 15.42 & 28.86 & 39.29 \\
2.0 & 9.6 & 0.85 & 0.59 & 16.11 & 15.46 & 28.92 & 39.36 \\
2.1 & 5.7 & 1.11 & 0.55 & 17.75 & 16.95 & 32.17 & 43.98 \\
2.1 & 6.4 & 1.14 & 0.59 & 17.75 & 17.00 & 31.90 & 43.46 \\
2.1 & 7.2 & 1.21 & 0.60 & 17.75 & 16.91 & 31.77 & 43.30 \\
2.1 & 7.8 & 1.21 & 0.59 & 17.75 & 17.06 & 31.53 & 42.75 \\
2.1 & 8.3 & 1.28 & 0.62 & 17.75 & 15.38 & 30.87 & 42.90 \\
2.1 & 8.8 & 1.34 & 0.63 & 17.75 & 17.75 & 31.20 & 41.63 \\
2.2 & 5.1 & 2.30 & 1.07 & 19.48 & 19.22 & 33.91 & 45.30 \\
2.2 & 5.8 & 2.40 & 1.12 & 19.48 & 19.16 & 33.93 & 45.38 \\
2.2 & 6.5 & 2.45 & 1.08 & 19.48 & 19.07 & 33.78 & 45.20 \\
2.2 & 7.2 & 2.68 & 1.14 & 19.48 & 19.07 & 33.75 & 45.13 \\
2.2 & 7.9 & 2.80 & 1.13 & 19.48 & 19.10 & 34.00 & 45.56 \\
2.2 & 8.5 & 2.94 & 1.11 & 19.48 & 19.03 & 33.66 & 45.01 \\
2.3 & 4.9 & 3.71 & 0.81 & 21.11 & 21.22 & 34.14 & 44.16 \\
2.3 & 5.4 & 3.75 & 0.79 & 21.11 & 21.50 & 34.28 & 44.19 \\
2.3 & 6.0 & 3.72 & 0.81 & 21.11 & 21.44 & 34.25 & 44.19 \\
2.3 & 6.6 & 3.78 & 0.77 & 21.11 & 21.53 & 34.26 & 44.13 \\
2.3 & 7.2 & 3.81 & 0.77 & 21.11 & 21.54 & 34.35 & 44.28 \\
2.3 & 7.7 & 3.77 & 0.81 & 21.11 & 20.54 & 33.96 & 44.37 \\
2.4 & 4.4 & 4.05 & 0.67 & 23.18 & 25.04 & 35.57 & 43.74 \\
2.4 & 4.8 & 3.93 & 0.70 & 23.18 & 24.45 & 35.32 & 43.77 \\
2.4 & 5.2 & 3.76 & 0.80 & 23.18 & 24.47 & 35.27 & 43.65 \\
2.4 & 6.1 & 4.00 & 0.70 & 23.18 & 24.42 & 35.63 & 44.32 \\
2.4 & 6.6 & 4.06 & 0.70 & 23.18 & 24.70 & 35.29 & 43.51 \\
2.4 & 7.2 & 4.15 & 0.66 & 23.18 & 24.26 & 35.07 & 43.46 \\
2.5 & 4.1 & 4.68 & 0.51 & 25.15 & 26.87 & 35.68 & 42.51 \\
2.5 & 4.6 & 4.95 & 0.45 & 25.15 & 27.96 & 36.18 & 42.56 \\
2.5 & 5.2 & 5.15 & 0.37 & 25.15 & 26.89 & 35.66 & 42.46 \\
2.5 & 5.8 & 5.38 & 0.32 & 25.15 & 28.06 & 36.13 & 42.41 \\
2.5 & 6.4 & 5.30 & 0.36 & 25.15 & 27.29 & 35.82 & 42.44 \\
2.5 & 7.5 & 5.46 & 0.30 & 25.15 & 26.81 & 35.63 & 42.48 \\
2.6 & 3.3 & 7.16 & 0.20 & 27.20 & 30.22 & 36.46 & 41.31 \\
2.6 & 3.7 & 7.18 & 0.19 & 27.20 & 31.08 & 36.76 & 41.21 \\
2.6 & 4.0 & 6.85 & 0.24 & 27.20 & 31.45 & 37.09 & 41.48 \\
2.6 & 4.3 & 6.79 & 0.22 & 27.20 & 31.64 & 37.23 & 41.58 \\
2.6 & 4.7 & 6.43 & 0.24 & 27.20 & 30.97 & 37.06 & 41.81 \\
2.6 & 5.4 & 6.46 & 0.23 & 27.20 & 30.06 & 36.63 & 41.74 \\
2.7 & 3.2 & 8.72 & 0.46 & 29.34 & 35.29 & 37.93 & 40.02 \\
2.7 & 3.4 & 7.79 & 0.42 & 29.34 & 33.98 & 37.76 & 40.76 \\
2.7 & 3.6 & 8.35 & 0.34 & 29.34 & 34.89 & 37.90 & 40.33 \\
2.7 & 4.2 & 8.67 & 0.25 & 29.34 & 34.03 & 37.36 & 40.03 \\
2.7 & 4.9 & 8.38 & 0.30 & 29.34 & 44.46 & 38.58 & 40.66 \\
2.7 & 5.2 & 8.46 & 0.16 & 29.34 & 30.33 & 36.01 & 40.45 \\
  \hline
  \caption{All the data used for generation of Figs. \ref{wavelength} and \ref{fig_damping}.}
  \label{all_data_table}
\end{longtable}

\section{Appendix C: supplementary figures}

In this Appendix, we present figures that help better illustrate some of the points of the paper, but that have some level of redundancy with the main body of the text.

\subsection{Independence of the damping coefficient on the wave amplitude}

\begin{figure}
  \begin{center}
    \includegraphics[width=.65\textwidth]{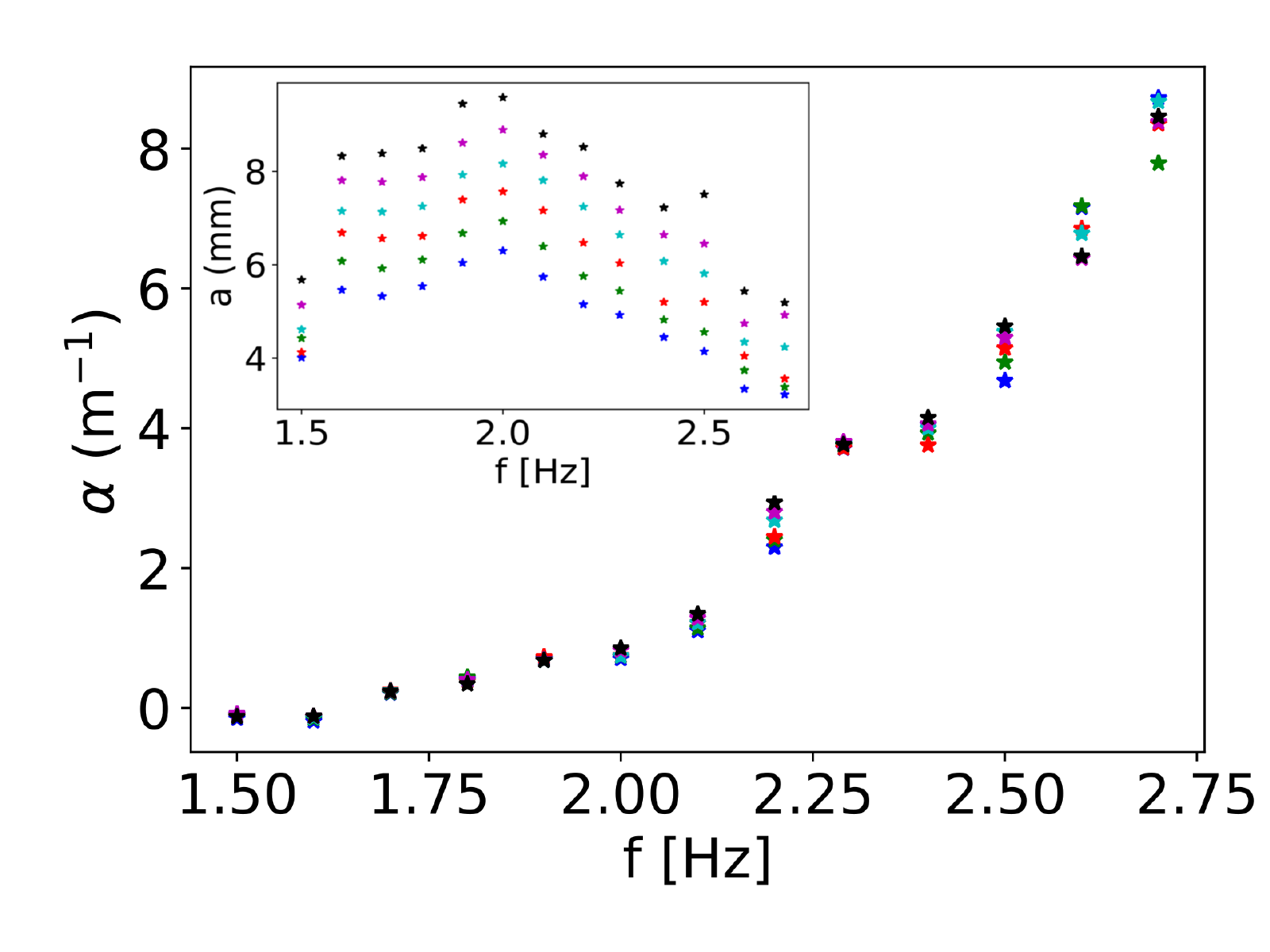}
    \caption{\label{non_dependence_alpha} Damping coefficient $\alpha$ as a function of the incoming wave frequency, plotted for each individual experiment (colored symbols). The color of the symbols indicates the wave amplitude used at each frequency, following the inset figure. We find that the damping coefficient has no significant trend when considering the influence of the wave amplitude.}
  \end{center}
\end{figure}

As stated in section 3.1, the damping coefficient is found to be independent of the wave amplitude in the linear regime that we investigate. This information is visible in Fig. \ref{fig_damping} as the 95\% confidence intervals are small, and also in the Table \ref{all_data_table} of Appendix B. However, to make this point clearer, we present a third illustration of this point in Fig. \ref{non_dependence_alpha}. As visible in Fig. \ref{non_dependence_alpha}, the wave amplitude does not influence the wave damping coefficient in any significant way (besides experimental uncertainties) for the range of wave amplitudes explored.

\subsection{Comparison of the damping coefficient with a power law}

\begin{figure}
  \begin{center}
    \includegraphics[width=.65\textwidth]{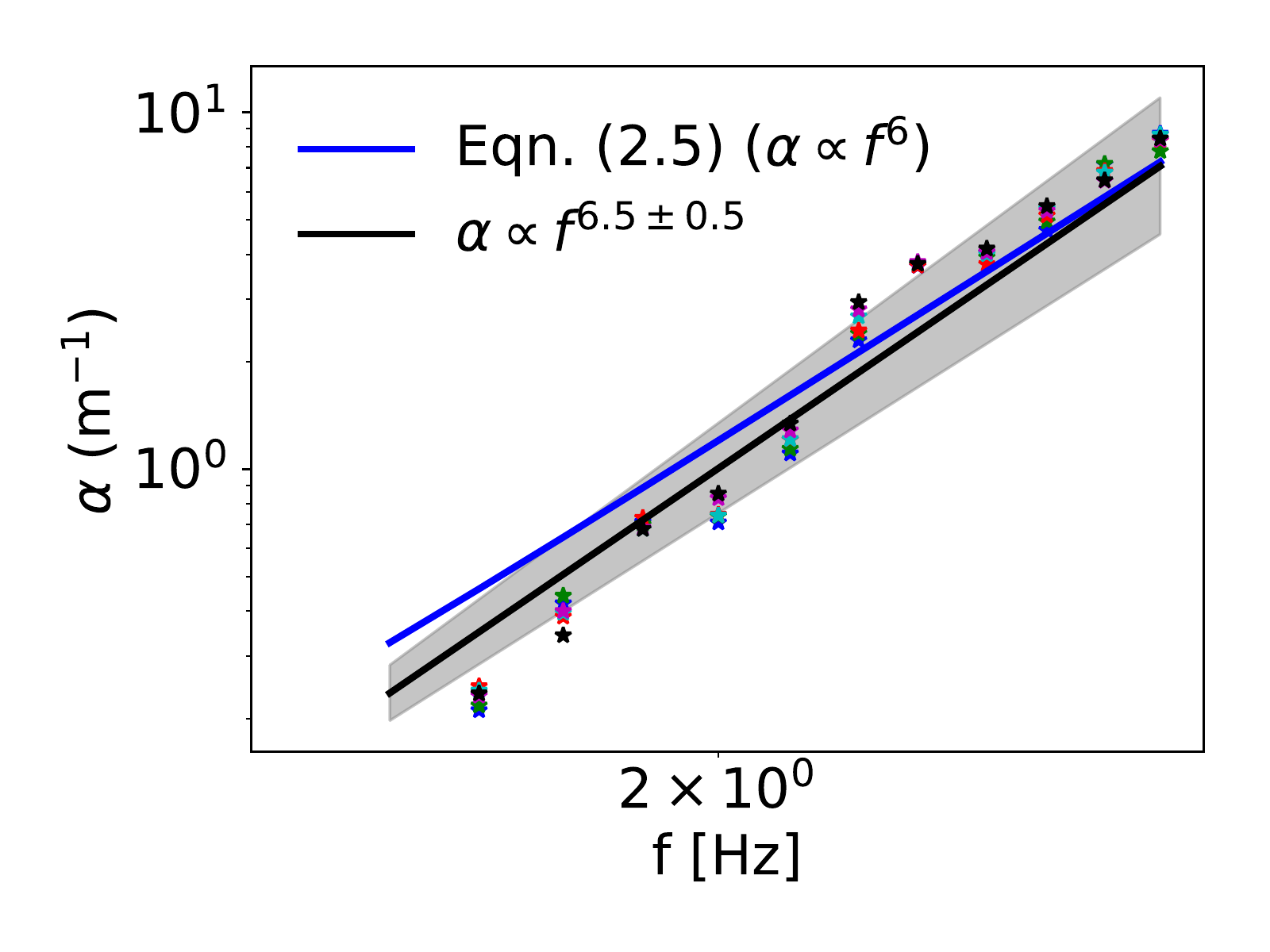}
    \caption{\label{alpha_power} Damping coefficient $\alpha$ compared with a power law scaling. The grayed area indicates the 3-$\sigma$ confidence interval for the exponent following a least-square fit. A good agreement with Eqn. (\ref{sutherland_damping_v2}) is obtained. We use the same color code for the individual experiments at each frequency as in Fig. \ref{non_dependence_alpha}.}
  \end{center}
\end{figure}

The frequency-dependent damping coefficient $\alpha$ is compared to a power law scaling, and results are presented in Fig. \ref{alpha_power}. Given the uncertainties introduced by the limited frequency range, a good agreement with Eqn. (\ref{sutherland_damping_v2}) is obtained.

\end{document}